# PHYSICS OF COMPLEX SYSTEMS: DISCOVERY IN AGE OF GÖDEL

Dragutin T. Mihailović, Darko Kapor, Siniša Crvenković, and Anja Mihailović

# Contents

*Preface*

Chapters 6-10 will be completed

**Chapter 6**

Kolmogorov and change complexity and their use in physics

**Chapter 7**

Separation of scales in complex systems

**Chapter 8**

Randomness representation of turbulence in fluids

**Chapter 9**

Physics of complex systems and painting

**Chapter 10**

Modelling the biophysical complex systems

**APPENDIX A**
Kolmogorov complexity and its derivatives

**APPENDIX B**
Change complexity

*Preface*

This book is the result of a joint venture of two physicists (also involved in meteorology and environmental studies) and two mathematicians (one algebraist and one in information technology field). Three of us are retired university professors, while the fourth one is a Ph.D. student. The inspiration for the book was found in the problems we had encountered in teaching and research, which required explanations that could not be given neither in the lectures nor in the research papers. So, the aim of the book is to present our considerations and experience to a broad audience and invite readers to further investigate various topics by themselves.

Why did we choose this title and content of the book? The first reason is encompassed by John von Neumann's comments: "... there have been within the experience of people now living at least three serious crises... There have been two such crises in physics---namely, the conceptual soul searching connected with the discovery of relativity and the conceptual difficulties connected with discoveries in quantum theory... The third crisis was in mathematics. It was a very serious conceptual crisis, dealing with rigor and the proper way to carry out a correct mathematical proof. In view of the earlier notions of the absolute rigor of mathematics, it is surprising that such a thing could have happened, and even more surprising that it could have happened in these latter days when miracles are not supposed to take place. Yet it did happen." The second reason was Kurt Gödel's original proof of incompleteness theorems with a paradoxical assertion that is true but not provable within the usual formalizations of number theory. In our opinion, his contributions to foundation of mathematics, philosophy, and physics unite the past, the present, and the far future into "Gödel's age." Gödel was an unique scientist and mathematician whose discovery was like "Deus ex machina." He "demolished" the old house and built a new one. Physics of complex systems was a fruitful field for traveling through "Gödel's age."

Serbian filmmaker Emir Kusturica states: "Today, everyone writes but no one reads." The same can be said for the scientific and mathematical community. We tried not to follow this already obvious trend. From the very beginning, authors were sure about what this book was not intended to be. First, it was not supposed to be a textbook, so scholarly explanations were avoided although they appeared sometimes when we thought that the majority of readers might not be familiar with all the concepts and terminology. Next, it was not considered to be an encyclopedia even though a wide range of topics was covered in the book. We used quotations and citing literature (around 10–20 percent of the chapter size) to keep the chapter sizes within a tolerable limit.

The creation of this book was an interactive process. There was first an interaction between authors to determine the main subjects of the book and its style. At the next stage, authors exchanged ideas on every particular subject, which was a very vivid interaction indeed. Once a consensus was achieved, the material was written down, and further steps were defined. The fourth author took care of consistency and readability of the book. The interaction with the editors who ensured that the book keeps a proper form and remains within the projected scope also contributed to the final form of the text.

The choice of examples was dictated by authors' experience, in particular by their research. Luckily, our research interests are broad, so there is a lot of different material in the book. One way of testing whether we were on the right track was to put the material on the internet (arXiv) when it was completed. Feedback helped us a lot.

For the end of the short history of this book, we left one detail that we noticed when half of the book was already written. In many examples, difference logistic equation appeared with an unknown but unexpected trait that was completely distant from our intuition. Its appearance reminded us of a picture in which the most magnificent animal on the Earth, blue whale, emerges from the most powerful part of the Earth and spins back into the mysterious depth of the ocean.

*Chapter 1*. This chapter is a discursive introduction to the book. We consider the current state of physics. We perceive further progress of physics as a synthesis of collective efforts that includes seemingly unrelated fields—physics, mathematics, chemistry, biology, medicine, psychology, and even arts. In that synergy, physics contributes to the domain of fundamental discoveries analogously to the way entities in a complex system contribute to the system as a whole. We support our standpoint through the consideration of complex systems, relationship between physics and mathematics, and increasing power of computational physics. *Chapter 2*. Great mathematician Kurt Gödel wanted to study physics at first. He wrote two papers, inspired by Einstein's theory of relativity, in which he derived a model of the universe allowing time travel. Einstein called this "a great achievement for the theory of relativity." Gödel's incompleteness theorems have exerted a strong influence not only on mathematics but also on physics. In this chapter, we explain his scientific background by analyzing his biography, give a concise proof of Gödel's incompleteness theorems, point out real possibilities and misconceptions about the application of these theorems in physics. In addition, we present his contributions to physics and information science. *Chapter 3*. One of the crucial philosophical issues is the question of time. In this chapter, we first offer (in no particular order) a list of the most important issues that are under discussion in the philosophical community regarding time. Then we shortly outline the understanding of time in theoretical and computational physics since in complex systems it operates concurrently at different scales, runs at multiple rates, and integrates the function of numerous connected systems (complex time). Finally, we discuss the conditions leading to the possible quantization of time. *Chapter 4.* This chapter deals with models and their use in computer simulation in the physical science, inherently involving many epistemological questions. Although heuristic considerations often overrule the problems of epistemology, it is sometimes necessary to make basic epistemological choices, especially in modeling. Being aware of its weak spots, physicists often ignore nonlinearity of phenomena and processes by applying the method of linearization. However, if we consider the existing nonlinearities in the object that we model as much as possible, we can recognize the following key points: (1) model choice, (2) continuous-time versus discrete-time in building the model, (3) model predictability (Lyapunov time), and (4) chaos in environmental interfaces in climate models. *Chapter 5*. In this chapter, we discuss information and its relation to physics. We address the following aspects: (1) physicality and abstractness of information, (2) the concept of information, (3) metaphysics of chance (probability), (4) information and no event registered, and (5) information in complex systems. *Chapter 6.* We elaborate on two information measures (Kolmogorov complexity and its derivatives as well as change complexity), which are useful in the analysis of various time series. The examples included here are the search for the patterns in analysis of Bell's experiments, identification of gravitational waves, and environmental fluid flows. The mathematical backgrounds of measures are presented in Appendix A and Appendix B. *Chapter 7.* At the beginning of this chapter, we set one view on separation scales in complex systems as a reflection of Gödel's incompleteness theorems. We point out the limits of renormalization group related to separation of scales. A need for new mathematics for scaling in complex systems is emphasized. At the end, we give some examples. *Chapter 8*. We discuss randomness in turbulent flows and its quantification via complexity and consider information measures suitable for its description. Finally, we consider connection of complexity of coherent structures with the turbulent eddies' size. *Chapter 9.* We make an elaboration about dualism between physics and art. We emphasize the place of physics of complex systems in creation of an impression about picture through 1. perception analyzed with change complexity; 2. recognition of order and disorder with entropy.

*Chapter 10.* In this chapter, we present some of the contributions of physics of complex systems to medical science. It is completed in terms of models and approaches that deal with intercellular communication, autoimmune diseases, and spatio-temporal brain dynamics.

The preparation of the text (both a creative and technical part) was performed under the permanent supervision of Ms. Betsy Byers, Editorial Assistant for Physics, CRC Routledge, Taylor and Francis, who was there for us whenever we needed advice, and we are grateful for her assistance. Prof. Zagorka Lozanov-Crvenković offered us help with parts of mathematics that we were not familiar with and also with the best way of presenting it, so we owe her a lot. The authors appreciate the contribution of Mr. Miloš Kapor who produced the artwork (original illustrations).

# Chapter 1

## Prolegomenon

### 1.1. Generality of physics

Looking at the current progress in science, technology, and its trends, one might be assured that discoveries and advances in the complex systems theory and applications will dominate the twenty-first century. This branch will improve our understanding of physical, biological, ecological, and social phenomena. Accelerated research that has occurred owing to the help of technology and a vast amount of data has led to new fundamental theoretical questions in science. These new theoretical problems cannot be solved with the help of traditional methods and require the development of a new interdisciplinary science of complex systems. Physics will be one of the most important parts since solutions to many problems in other sciences rely on physical laws. The contribution of physics to modern science should be defined carefully because its universality and wide applications resulted in the generality of physics. The generality of physics is an idea that phenomena (irrespective of their origins and nature) fall under the physics' domain and should be explained by its laws. This is especially accepted in biology, but to which extent physics explains biological complex systems.

According to many dictionaries, physics is concerned with nature and properties of matter and energy. It includes mechanics, heat, light and other radiation, sound, electricity, magnetism, and structure of a substance. The concept of matter often equates with a material substance that constitutes the observable universe and together with energy, which appears also in the form of fields, forms the basis of all objective phenomena. *Inert matter* comprises matter that is not a seed, including broken seeds, sterile florets, chaff, fungus bodies, and stones. A remarkable part of the physical community takes it for granted that inertia is intrinsic to matter although inertia only refers to the matter in the universe according to Ernst Mach, a German physicist. On the other hand, living organisms are structured, which means that they contain specialized, coordinated parts. They are made up of either one cell (unicellular organisms) or more cells (multicellular organisms) that are the fundamental units of life. It appears that pioneers in exploring these two complex worlds are physics and biology. Nowadays physics is considered general, and biology is regarded as merely particular by many scientists, but this assertion is improper. It can be said that physics incorporates all matter in nature including organisms since they belong to material nature. From such an ideal perspective, biology is really a part of physics (when we say physics, we mean *contemporary* physics). Still, physical laws are applicable to the limited number and types of material systems; physics is inherently inadequate to accommodate phenomena in biology. The author of the book *What Is Life* [1] Erwin Schrödinger was one of the outstanding theoretical physicists in the twentieth century, perhaps in a previous millennium. He regards physics as the ultimate science of material nature and concludes that organisms are repositories of what he calls new physics. Robert Rosen [2] says that Erwin Schrödinger while permanently asserting the universality of contemporary physics, also points out the complete failure of its laws to explain anything significant about biosphere and its structure. Albert Einstein also states the same in a letter sent to Leo Szilard but describes that more vividly: "One can best feel in dealing with living things how primitive physics still is" [3]. By contrast, Jacques Monod [4], three decades after the appearance of Schrödinger's essay, writes, "Biology is marginal because—the living world constituting but a tiny and very 'special' part of the universe—it does not seem likely that the study of living things will ever uncover general laws applicable outside the biosphere" [2]. Therefore, in this book we only address the breakthroughs of physics in biology and medicine for which physical processes are obvious and clearly defined.

What is the level of generality of physics or any other scientific or mathematical discipline? We can assume intuitively that "the level of generality" of a theory characterizes the class of situations that the theory can deal with or to which the theory can accommodate. How can something like that be measured? It is illustrative to look at number theory comprising many conjectures that no one has ever been able to prove or disprove. The rising question was whether number theory was general enough to explain problems that had arisen. The situation became more interesting when Kurt Gödel showed how to represent the assertion about number theory within number theory [5]. Considering things from this perspective, he was able to demonstrate that for any given set of axioms in number theory, there are always propositions that are similar to theorems but cannot be proved from these axioms. If such a situation already exists in number theory, then we can imagine how it is difficult to consider a similar problem in physics. Gonzalo [6] explains that Stephen Hawking [7] believes that Gödel's incompleteness theorems [2] make the search for the theory of everything impossible. He reasons that because there exist mathematical results that cannot be proved, there exists physical results that cannot be proved as well. Exactly how valid is his reasoning? Opinions on this view have been very divergent. We support the opinion that Stephen Hawking [7] does not claim that results are unprovable in an abstract sense. He assumes that the theory of everything will be a particular finite set of rules and presents an argument that no such set of rules will be sufficient. Gonzalo states that "the final statement that it may not be possible to formulate a theory of the universe in a finite number of statements, which is reminiscent of Gödel's theorem" [6].

The above question is exactly the question raised by reductionism in physics that is understood as methodological reductionism. It is the attempt to reduce explanations to smaller constituents and to explain phenomena as relations between more fundamental entities [8] that connect theories and knowledge within physics. Note that the problems associated with physical reductionism, at least applied naively, are: 1. It misses the emergent properties of a system (reductionism assumes that emergent properties are nothing more than the sum of the reduced properties applied over a very large scale). 2. It is misapplied in biology and other sciences [9]. The question we talk about is an assertion, conjecture, or belief that pertains to the generality of contemporary physics itself. Unlike Goldbach's conjecture in number theory, conjectures in the physical world are not based on any direct evidence. "It is rather indirect (circumstantial) evidence, insofar as evidence is adduced at all. In short, it rests on faith" [2]. This faith was described metaphorically by the Serbian writer Borislav Pekić in the novel *Atlantis* [10]: "The existence of spirits in principle does not contradict any law of physics. It is in contrast to the mind that civilization has modeled on empirical evidence for centuries." Conjectures are based on general experience and very limited since they are always subject to the so-called "black swan" effect. "Black swan" is an expression for the event that is very rare, completely unexpected, and unpredictable. It has a large impact since it can easily cause the collapse of a complete scientific structure because of conjectures that excluded its existence [11]. At the end of this subchapter, we must stress that everything we said about physical systems was quite general. We later make a clear distinction between the systems with the infinite or very large number of constituents and the systems with interfaces compared to the rest of physical systems.

In order to offer a better insight, let us just summarize perspectives. Maybe a right direction is to list the problems that will be in the focus of the scientific community in the twenty-first century. Some possible choices, without hierarchical order, are: (1) whether we are alone in the universe, (2) the emission of harmful greenhouse gasses and climate change, (3) the understanding of consciousness, (4) decision-making in an insecure world, (5) the extension of both the maximum and average lifespan, (6) whether the culture is only characteristic of people, (7) managing the Earth resources; (8) the importance of the internet, (9) the use of master cells in the future, (10) the importance of maintaining biodiversity, (11) the role of reengineering and climate change, and (12) the importance of new vaccines. Let us also mention that, during the preparation of this book, we submitted some early versions of chapters, and interested reader can look at an extended version of this subchapter [12].

## 1.2. Physics: A crisis that has been lasting for a century! Is that really so?

Many articles on the crisis in science have been published starting from the pioneering paper written by Freistadt [13]. Nonetheless, the clear identification of its origins remains elusive to most commentators [14]. Because of the increasing public impact of the crisis on trust in institutions, authors argue that 1. the crisis in science (hence physics) exists and includes positions and social functions of science (crisis context); 2. the mainstream interpretation of its causes is not sufficient and has to be complemented with insights provided by some researchers who predicted the current dilemma (a lack of root causes); 3. the profound transformation of society and impact of science on society are, without doubt, induced by this crisis (scientists have also contributed to its creation and had a great influence on preserving the status quo [science vs stake context]); 4. there are some social mechanisms that can be applied to improve this situation including the important changes in behavior and social activity of the scientific community (social context) (emphasis added).

The crisis and its consequences are clearly reflected in awarding the Nobel Prize. Fortunato [15] points out that the time between publishing research and receiving the Nobel Prize has become longer. This trend is the least present in physiology or medicine and the most present in physics. Prior to 1940, 11 percent of prizes in physics, 15 percent of prizes in chemistry, and 24 percent of prizes in physiology or in medicine were awarded for research more than 20 years old. Since 1985, those percentages have risen to 60 percent, 52 percent, and 45 percent, respectively.

"The crisis of physics" implies that everything is affected and that all results are questionable. According to our opinion, this is not the issue, and we just state that the crisis exists within physics, or in physics. What is the crisis in physics? We perceive it as the absence of a fundamental discovery, or more precisely, the crisis in physics arises and lasts until a new fundamental discovery is made. This expectation is not the expectation of a person who will never appear similar to the one in Samuel Beckett's tragicomedy *Waiting for Godot* [16]. Many modern concepts about the crisis in physics envisage ending identical to the end of this tragicomedy; we do not see ending but rather transition. From our standpoint, a fundamental discovery is a discovery made by scientists who possess the exceptional abilities of thinking and revolutionize physics. We label such progress as *vertical* progress, and it does not matter if it involves a micro, macro, or mega physical world. Other discoveries in physics, the confirmation of existing hypotheses or experimental results providing or needing explanations, are entitled *horizontal* progress. We must admit that our definition of the fundamental discovery is not complete since it assumes a large amount of horizontal progress that precedes the discovery. For this reason, we are not confident whether the experimental proof of the expanding universe and the quark hypothesis should be treated as fundamental. In our opinion, the three most recent fundamental discoveries in physics are: 1. Quantum Theory (Max Planck, 1900) that introduces an individual unit of energy or quant;

the sources are the two outstanding communications of Plank to the Berlin Academy (extraordinary fit of the radiation formula and statistical justification by introduction of the discrete energy elements) collected in the paper by Boya [17]. 2. Theory of special relativity introduced by Albert Einstein in 1905 that was his annus mirabilis year when he published three papers [18]–[20] in *Annalen der Physik*. These papers in which he revolutionized the concepts of space, time, mass, and energy made major contributions to the foundation of modern physics. 3. Theory of general relativity (Albert Einstein, 1915) that is concerned with macroscopic behavior and describes large-scale physical phenomena. Sauer [21] provides the first comprehensive overview of the final version of the general theory of relativity published by Einstein in 1916 after several expositions of the preliminary versions and their latest revisions in 1915. These theories treat a micro world (molecules, atoms, and elementary particles), macro world (our planet and people), and mega world or the universe (stars, planets, etc.).

It should be stressed that almost all facts relevant to fundamental discoveries were previously available to scientists. Planck was attempting to explain his extraordinary fit. Before the theory of general relativity, non-Euclidean geometry was already well-known owing to the research of Lobachevsky, Riemann, and Minkowski. The examples of findings prior to the theory of special relativity are: 1. The formulation of the *Lorentz force* as the force acting on charges moving in the magnetic field, which was essential for interactions between the currents. 2. *Lorentz transformations* that are considered constitutive for the theory of special relativity. They were invented by the efforts of Woldemar Voigt, George FitzGerald, and Hendrik Lorentz, while their derivation was described by Heras [23]. 3. The form of *Einstein's equation that relates mass to energy* was known to Henri Poincaré [24] who stated that if one required that the momentum of any particle that is present in an electromagnetic field plus the momentum of the field itself should be conserved together, then Poynting's theorem [25] predicted that the field acted as a "fictitious fluid" with mass satisfying the expression Einstein used later. Unfortunately, the whole theory was based on the concept of ether. The only difference between the above-mentioned scientists and Einstein is the way they interpreted available facts. His interpretation made his discovery fundamental and raised physics to a higher level.

What has caused and created this crisis? It seems that the crisis in physics has at least three aspects: 1. *Psychological aspect*. Let us make a digression regarding the term "escape" in the context of positive freedom defined by Fromm [26]. Based on Fromm's definition, it is the capacity for "spontaneous relationship to man and nature, a relationship that connects the individual with the world without eliminating his individuality" [26]. Freedom is also accompanied by loneness as well as an inability to exert individual power, and "we use several different techniques to alleviate the anxiety associated with our perception of freedom, including automaton, conformity, authoritarianism, destructiveness, and individuation" [27]. The most common of these mechanisms is conformity. Fromm states that people conform to larger society and gain power and a sense of belonging by behaving similarly. This power of the masses assists us in not feeling lonely and helpless, but unfortunately, it removes our individuality. How is this phenomenon connected with the crisis in physics? Specifically, the escape from science can be seen as an analogy to the escape from freedom or to the "escape from freedom of choice of the research subject" [26]. How is it possible for physicists to escape from physics today? Physicists are willing to believe that cause can be determined, and they accept an obvious solution and declare it a right explanation. On the contrary, they avoid a random event that almost always affects and changes physical systems. Why? Because physicists find nearly impossible to deal with random events that evade clear explanations and rules. This ignorance of chance in physical processes leads to scientific conformity and creation of mainstream in science (hence physics), or the mechanism of escape from physics. Note that the syntagma mainstream refers to physicists who have adopted attitudes that are opposite to Anderson's understanding of physics. Finally, our mind is filled with causality, and we primarily say because instead of accepting a random event. 2. *Epistemological aspect.* a. *Limits of the precision of certainty and strength of Gödel's theorem*. There is no doubt that physics is approaching some limits; the only question is whether they are the limits to how far physics can reach or its end. To clarify our position, we rely on Gödel's incompleteness theorems. Barrow [28] considers some informal aspects of these theorems and their underlying assumptions and discusses some of the responses to these theorems by those seeking to draw conclusions from them about the completability of theories of physics. In the same paper, he states "that there is no reason to expect Gödel incompleteness to handicap the search for a description of the laws of Nature, but we do expect it to limit what we can predict about the outcomes of those laws." Stanley Jaki [29] believes that Gödel's theorem prevents us from gaining the understanding of the cosmos as a necessary truth: "Clearly then no scientific cosmology, which of necessity must be highly mathematical, can have its proof of consistency within itself as far as mathematics goes. In the absence of such consistency, all mathematical models, all theories of elementary particles, including the theory of truth that the world can only be what it is and nothing else. This is true even if the theory happened to account with perfect accuracy for all phenomena of the physical world known at a particular time." It constitutes a fundamental barrier to the understanding of the universe: "It seems that on the strength of Gödel's theorem that the ultimate foundations of the bold symbolic constructions of mathematical physics will remain embedded forever in that deeper level of thinking characterized both by the wisdom and by the haziness of analogies and intuitions. For the speculative physicist this implies that there are limits to the precision of certainty, that even in the pure thinking of theoretical physics there is a boundary…An integral part of this boundary is the scientist himself, as a thinker" [30] (emphasis added). b. *Limits of decoding information*. Advances in information theory are of critical importance in modern physics. For instance, they are important for detecting the gravitational waves data provided by LIGO (Laser Interferometer Gravitational-Wave Observatory) [31]. However, even if we move the borders of cognition in physics, it remains the problem of decoding information. Each meaningful information is encoded in patterns that have different complexity and is always less complex than maximum complexity (randomness). In contrast to dealing with easily observed and immediately decodable patterns (e.g., visual input, native language), our cognitive capacity struggles extracting information from noise. Furthermore, the presence of a pattern (structural information) in noise is a necessary but not a sufficient condition for the presence of the meaning (semantic information). Besides, there are other aspects that should be considered: 1. How we interpret information? 2. Can we search for information outside established concepts? Namely, people are not used to thinking abstractly [11], so they need confirmed rules and facts for further induction. Consequently, most scientists are focused on and seek the confirmation of what happened and ignore what could have happened. We lack the dimension of abstract thinking, and we are not even aware of that. Therefore, physicists initiate "the trap of induction" and search for information that is easy to comprehend and not information that is hardly reachable to our minds.

The effect of scientists' emotions and intuition on science and scientific work is worth mentioning. Intuition is the accumulation of experiences that are assimilated unconsciously [32]. Einstein [33] writes an interesting consideration on the importance of intuition: "I believe in intuition and inspiration. Imagination is more important than knowledge. For knowledge is limited, whereas imagination embraces the entire world, stimulating progress, giving birth to evolution. It is, strictly speaking, a real factor in scientific research." Emotions are essential in science, including the justification as well as the discovery of hypotheses [34]. Note that Gödel's work has had a huge impact on the speculations about the limitations of the human mind [28]. 3. *Social and economic aspects.* The fact that a major breakthrough in physics has not recently appeared influences the general social attitude towards research. Building large machines (colliders, telescopes) cannot be justified neither by pure curiosity nor by promising results, but mostly, just as a cosmic flight or Formula 1 racing contribute to the technology, by immediate applications in everyday life. This reasoning has resulted in misinterpretations of scientific results just to make them look like something applicable. This aspect is sometimes forgotten, but when it comes to experimental research, it is of great importance.

How do we see "the ending of physics"? In *The End of Physics* [35], David Lindley argues that 1. the theory of everything derived from particle physics will be full of untested and untestable assumptions; 2. if physics is the source of such speculation, it will eventually detach from science and become modern mythology. This will be the end of physics as we know it (emphasis added). Indeed, all big problems in physics have been solved or soon will be. Apparently, there are not more truly fundamental discoveries, such as quantum mechanics, relativity, the Big Bang and beyond (cosmology), left to be made. Therefore, some possibilities for future trends in physics are: 1. It will be aimed at less interesting activities (the research on detailed implications of basic theories, applied problems, etc.). 2. It will be less stimulating and attractive to stellar scientists who are able to make vertical progress. The Nobel Laureate Philip Anderson presents his observations on physics, its future progression, and directions towards fundamental discoveries [36]. He differentiates two trends in physics— "intensive" and "extensive" research. Intensive research includes fundamental laws, while extensive research comprises the explanation of phenomena in terms of established fundamental laws: "There are two dimensions to basic research. The frontier of science extends all along a long line from the newest and most modern intensive research, over the extensive research recently spawned by the intensive research of yesterday, to the broad and well-developed web of extensive research activities based on intensive research of past decades" [36]. He also points out the nature of the hierarchical structure of science and physics. Entirely new properties appear at each level of complexity, and the understanding of new behavior requires fundamental research. Anderson [36] describes his thoughts on fundamental science through his experience: "The effectiveness of this message [about intensive and extensive sciences] may be indicated by the fact that I heard it quoted recently by a leader in the field of materials science, who urged the participants at a meeting dedicated to 'fundamental problems in condensed matter physics' to accept that there were few or no such problems and that nothing was left but extensive sciences which he seemed to equate with device engineering" (emphasis added). His comments can be interpreted as the strong criticism of the efforts at replacing science with technology or even as an attempt to transform philosophy into science; philosophy is not science since it employs rational and logical analysis as well as conceptual clarification, while science employs empirical measurements. Similarly, technology cannot replace science. In our opinion, if we take into consideration his conclusions, we can notice continuity and not the end of physics. Our perspective on future developments is similar to Dejan Stojakovic's [37]: "Further progress of physics will require a synthesis of collective efforts in seemingly unrelated fields—physics, mathematics, chemistry, biology, medicine, psychology, and even arts such as literature, painting and music." An additional comment clarifies this sentence more precisely. Physics can contribute to the domain of fundamental discoveries analogously to the way entities in a complex system contribute to a whole system.

## 1.3. Complex systems in physics

A complex system consists of components whose properties and interactions create and influence the whole system. There are many concepts in the science of complex systems, but emergence and complexity are the most fundamental and important. *Emergence* is the complex system's behavior that is not regulated by the properties of its individual parts but by their interactions and relationships. That is to say, emergent behavior occurs only when individual parts interact. Jeffrey Goldstein [38] defines emergence as "the arising of novel and coherent structures, patterns and properties during the process of self-organization in complex systems." There are two notions related to emergence—weak and strong emergence. When interactions between entities at lower levels form new properties at higher levels, then this behavior of a complex system refers to *weak emergence* (usually called emergence). In the case of weak emergence, those low-level processes can be determined, at least approximately. It should be emphasized that "truths concerning that phenomenon are *unexpected* given the principles governing the low-level domain" [39]. Therefore, weak emerging properties are scale dependent—they are noticeable only if the system is large enough to exhibit a phenomenon. One nice thought on the connection between Gödel's incompleteness theorems and scalability in physics is described by Arthur Seldom in *The Oxford Murders* [40].

> That exactly the same kind of phenomenon occurred in mathematics, and that everything was, basically, a question of scale. The indeterminable propositions that Gödel had found must correspond to a subatomic world, of infinitesimal magnitudes, invisible to normal mathematics. The rest consisted in defining the right notion of scale. What I proved, basically, is that if a mathematical question can be formulated within the same "scale" as the axioms, it must belong to mathematicians' usual world and be possible to prove or refute.

But if writing it out requires a different scale, then it risks belonging to the world—submerged, infinitesimal, but latent in everything—of what can neither be proved nor refuted.

On the other hand, *strong emergence* is structurally unknown to us—that is, we cannot deduce high-level properties from low-level processes or laws. Undoubtedly, strong emergence involves some processes that we are unable to solve. The only example of strong emergence outside physics is consciousness, while quantum mechanics is a strong candidate for strong emergence within physics. Many researchers have recently challenged the interdependence of consciousness from quantum physics. Hypothetically, we can state that weak and strong emergence are the same with the one difference—our capability and incapability to find processes/laws governing the low-level systemic layer. Weak emergence can be simulated, and a crucial point is that the basic entities for the most part remain independent after computer simulation. If this does not happen, then a new entity with new emergent properties is created. In contrast, there is a broad consensus that strong emergence cannot be simulated nor analyzed. *Complexity* is a nontrivial regularity having its origin inside the system's structure. There is no unique explanation of complexity, and the most general definition is that the system exhibits complexity when its behavior cannot be easily explained by examining its components. Scientists use various complexity measures to assess complexity, especially in the analysis of time series since they are the only available evidence of the complex systems' nature.

We consider a *complex system* (fig. 1.1) as a collection of entities (circles). Each component interacts with others via simple local rules and the possibility of feedback (arrows). When they interact, a new feature arises (level 1). The complex system cannot be decomposed nontrivially into a set of basic entities for which it is the logical sum [2] (to have the character of an emergent phenomenon this new feature is completely unexpected [level 2]). The new property characterized by $(\zeta, P)$, where $\zeta$ represents data, and $P$ the probability distribution.

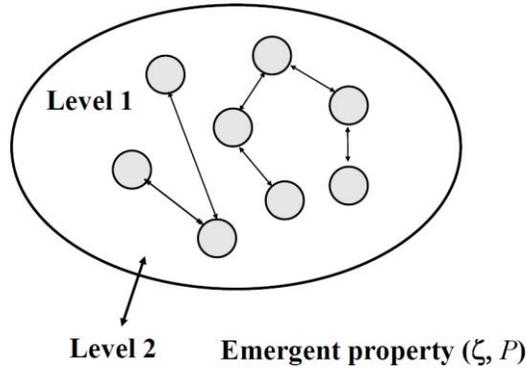

Figure 1.1. Towards the definition of a complex system and the concept of emergent property. (Reproduced by permission from [41].)

Most complicated systems in physics are not complex, but most complex systems in physics are complicated [42]. It is difficult for physicists to comprehend a complex system with various scopes and scales [36]. Paradoxes arise because we are often able to understand a part of the whole system. Further, we usually 1. embrace a microscopic or macroscopic level but not both simultaneously; 2. analyze either the system or the environment but not their mutual interaction. It seems that category theory is a promising choice for modeling a physical system that is hierarchically organized [43]. This was proposed as an alternative to the foundations of mathematics with the assumption that morphism is the basis from which everything is built up [44].

**1.4. Physics and mathematics walking together along a narrow path**

In ancient times and even in the Middle Ages, mathematics and physics were developing together, mostly as the parts of philosophy. A typical result of such a relationship is a vector that resulted from the study of forces (remember that the name vector is derived from the Latin verb *vehere* that means "to pull"). It was only later, particularly in the nineteenth century, that they became completely separated, and mathematics evolved into an abstract science. Scientists have explored the connection between mathematics and science for more than a century. Almost all papers on this topic begin with "the unreasonable effectiveness of mathematics"— the phrase created by the physicist Eugene Wigner [45] in 1960. He wonders how "the enormous usefulness of mathematics in the natural sciences is something bordering on the mysterious and that there is no rational explanation for this" [45]. We think that this reasoning was described much earlier in the famous book *The Aim and Structure of Physical Theory* [46] written by the French physicist Pierre Duhem. In the parts "The Aim of Physical Theory" and "Mathematical Deduction and Physical Theory", he explains that 1. theoretical physics is subordinated to metaphysics; 2. mathematical deductions are sometimes useful in physics, but axioms are usually self-evident truths from which theorems are derived deductively. On the contrary, physical axioms are not strongly self-evident, but physicists proceed with making new conclusions from those axioms even with mathematical deduction. His book was published when quantum theory and theory of special relativity were broadly known, while Albert Einstein submitted the theory of general relativity for publication in 1915 (in which he used Riemann's geometry—maybe the most brilliant use of mathematics in physics).

Mathematics is the science of order, rules, structure, and logical connections. It explores patterns and possible relationships between abstractions irrespective of whether they correspond to phenomena in physics. Gödel's incompleteness theorems basically express that a particular part of mathematics cannot be formalized. This does not necessarily mean that formalization would not be a strategy in the study of mathematical systems. These theorems rather indicate that the limits of formalization are the universal strategy. Mathematics provides a theory that scientists can recognize and even use to find similarities between two completely different structures in physics and other sciences. (Noam Chomsky used mathematics to "see" abstract patterns that we recognize as grammatical sentences.)

Research in theoretical physics consists of the following steps: 1. Scientists formulate a hypothesis usually without validating certain conditions. 2. They decide on mathematical methods for their hypothesis. Currently, there are several problems related to theoretical physics and its mathematical formalism. Prior to the choice of a mathematical model, there is a limit imposed by the nature of physics—physics does not include what happens in infinity because this is *beyond* its relative horizon. "The problem of philosophy is to acquire a consistency without losing the infinite into which thought plunges" [47]. This problem differs from the problem of explaining chaos with the condition of abandoning infinite movements and limitation of speed as *condition sine qua non*. Chaos is considered as a movement from one determination to another by physicists; from this perspective, chaos introduces disorder and unfastens every consistency in infinity. It seems that the problem is not the determination of chaotic behavior but the infinite speed at which elements are shaped and disappear. Further, physical theories are formalized throughout mathematical models (fig. 1.2) that can be undecidable; therefore, the problem of *undecidability* exists in physics, especially in condensed matter physics [48]. Last, mathematical models require a priori or complete knowledge about the conditions of their validity.

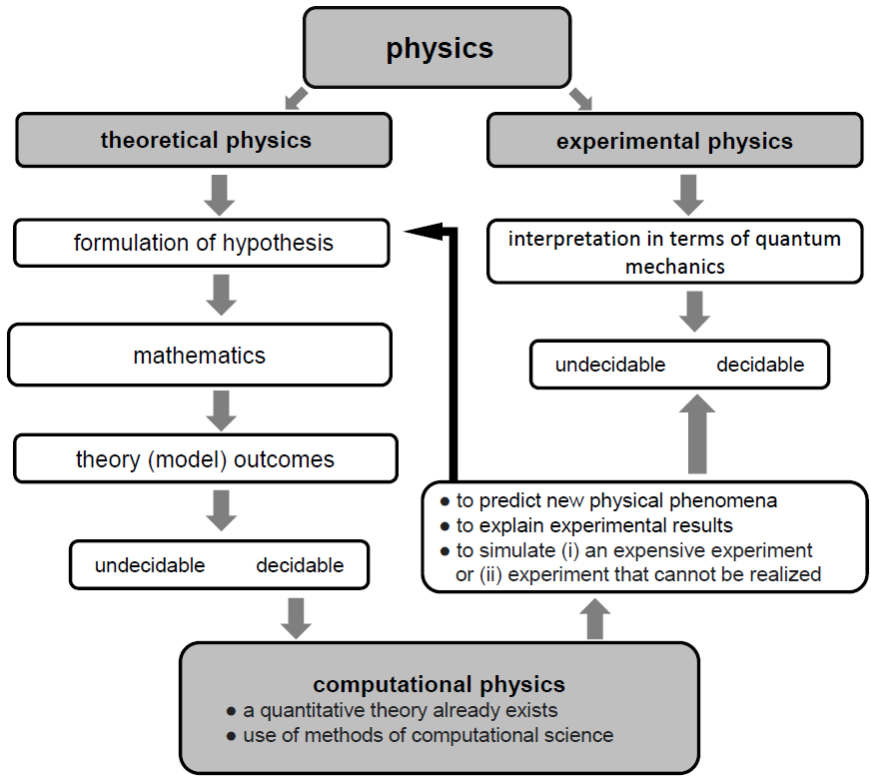

Figure 1.2. Place of mathematics in branches of physics.

Alex Harvey emphasizes that a physicist looks for an explanation by mapping physical phenomena onto mathematical objects. This must represent an isomorphism, and then mathematics can be applied [49]. Still, one must be very careful. Physicists often recognize established mathematical formalism, and it is rare that a certain mathematical discovery is inspired by physics. Variational calculus or strings might be considered as examples.

Mathematics is related not only to theoretical physics but also to experimental physics (fig. 1.2). Whether it acts as a bridge between them or addition to this classification is still a matter of debate. This applies also to computational physics because it connects two physical branches and is an independent discipline at the same time. Its foundation is theoretical physics, but it additionally implements and combines computer simulations and applied mathematics with physics to solve complex problems. As a result, this branch can help in predicting new phenomena, explaining experimental results, and replacing expensive experiments or experiments that cannot be conducted. The output of a numerical simulation (fig. 1.3) is the behavior of molecules in an ideal gas obtained by Arsenić and Krmar's model [50]. This relatively simple simulation illustrates 1. the spatial distribution of molecules of ideal gas after each collision; this cannot be determined experimentally: 2. the spatial and temporal distribution of ideal gas molecules; this is helpful in discovering a new physical phenomenon.

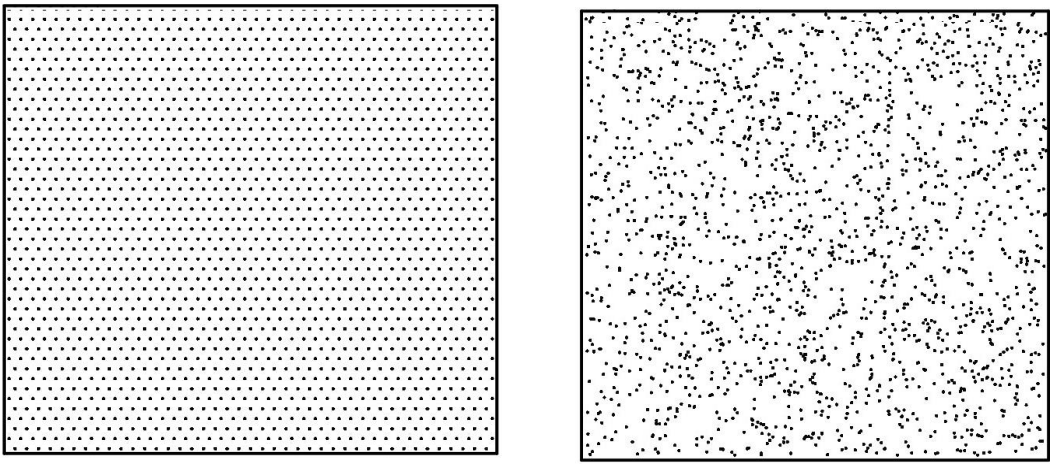

Figure 1.3. Spatial distribution of molecules in an ideal gas obtained by the numerical simulation at the beginning—initial conditions (*left*) and after one million collisions (*right*). (Figure courtesy of Ilija Arsenić and Miodrag Krmar**.**)

When we state that mathematical disciplines might be most helpful in physics in the future, we are aware that this assertion is not unambiguous. At this point, we focus on *category theory* because of its huge potential for future applications in physics. Category theory relates results from two different mathematical fields algebra and topology—the parts of foundations of mathematics. It has a central place in contemporary mathematics, theoretical computer science, theoretical physics, and other scientific disciplines. Let us imagine an abstract world that includes topological spaces and groups of algebraic topology, or the world of unspecified *objects*; this is the world of continuous mapping, group of *homeomorphisms* and also *morphisms* that are not specified. Our requirement is that those morphisms can be composed. This is completed with a *functor* whose role is to associate objects by morphisms while preserving the identities of structures. Let us suppose that we use category theory to create a model. Then "if we think of functorial images constituting *metaphors* for what they are image, we may ask whether two different such metaphors are the same or not" [2]. Terms "target" and "source" are commonly used to explain the metaphor comprehension; for example, in the metaphor " $T$ is $S$ " or "$T$ is like ", $T$ is a target, and $S$ is a source. In general, the target is moderately unknown, while the source is moderately known [51]. Then we compare relations and functors to assemble relations between them. Here is the point where modeling enters the picture explicitly not (up to now) between the source ($S$) and target ($T$) of a single functor but between targets of different functors defined on a common source. This type of comparison is called *natural transformation* that, in history of category, came the first, while the rest of category theory came as the reinforcement for it [2]. Thus, we say that we model $S$ in . This is of crucial importance for physics, particularly for the physics of complex systems, because this formalism provides an opportunity to find *relations* between objects. Usually, physicists perform the following steps that are not always productive work: they deal with objects by using selected mathematics and then explore potential relationship between them. As the final observation, let us pose a question: How should we look at the relationship between physics and Gödel's incompleteness theorems and his other ideas related to physics? Perhaps the most condensed description is given by John Barrow in the conclusion of his paper [28].

> Thus, in conclusion, we find that Gödel's ideas are still provoking new research programmes and unsuspected properties of the worlds of logical and physical reality. His incompleteness theorems should not be a drain on our enthusiasm to seek out and codify the laws of Nature: there is no reason for them to limit that search for the fundamental symmetries of Nature in any significant way. But, by contrast, in situations of sufficient complexity, we do expect to find that Gödel incompleteness places limits on our ability to use those laws to predict the future, carry out specific computations, or build algorithms: incompleteness besets the outcomes of very simple laws of Nature. Finally, if we study universes, then Gödel's impact will always be felt as we try to reconcile the simple local geometry of space and time with the extraordinary possibilities that its exotic global structure allows. Space-time structure defines what can be proved in a universe.

# Chapter 2

## Gödel's Incompleteness theorems and physics

**2.1. Gödel's biography and historical background of incompleteness theorems**

At the dawn of the twentieth century, David Hilbert (1862–1943), who posed so-called Hilbert's second problem, asked for a direct proof of the consistency of the theory of numbers, integers, or reals. In 1931 a shocking solution was found by Kurt Gödel who showed that the consistency of any theory containing the theory of numbers could not be proved within the theory itself. Hence, no theory that must be a foundation for mathematics can justify itself and must search for justification in an external system. Consistency means that, within the theory, we are not able to prove that both sentences $A$ and $\neg A$ are theorems of the theory. Therefore, no consistent theory containing the theory of numbers can be complete at the same time in the sense that all mathematical truths expressible in its language can be proved in the theory. One of the truths that cannot be proved is its own consistency. This result of Gödel is called *the incompleteness theorems.* In this subchapter, we describe the life and achievements of Kurt Gödel. It is obvious that almost all mathematical foundations lead to Hilbert—the real leader of mathematical science from the end of the nineteenth to the first half of the twentieth century. Hence, we first describe his work shortly.

Hilbert was born in Kaliningrad (Russian Federation). Apart from spending one semester at the University of Heidelberg, he obtained his mathematical training at the University of Königsberg at which the only full professor of mathematics was Heinrich Weber. Owing to his lectures, Hilbert became aware of number theory and function theory. At that time, he met the young mathematician Hermann Minkowski with whom he later cooperated fruitfully. The German mathematician Ferdinand Lindemann was invited to replace Weber. He suggested that Hilbert should research into a problem in the theory of algebraic invariants and was his thesis advisor. Ph.D. thesis was defended in 1885, and Hilbert was appointed the full professor in 1893. He solved famous Gordan's problem [1] (a problem in the theory of algebraic invariants) around 1888. Hilbert's revolutionary work was generally recognized and accepted afterward. Hilbert saw the number theory as "a building of rare beauty and harmony" [1]. At the Annual Meeting of the German Mathematical Society in 1893, Hilbert and Minkowski were selected to prepare a report on the current situation in the theory of numbers. It had been decided that the report should be divided into two parts. Minkowski would treat rational number theory, while Hilbert would deal with algebraic number theory. Hilbert became the world's leading mathematician. His interesting and fruitful activity was his discussion with Henri Poincaré on the reciprocal relationship existing between analysis and physics. This inspired him to present the list of unsolved problems in his major lecture at the II International Congress on Mathematics in Paris in 1900. Hilbert's dream was to, once for all, clarify the methods of mathematical reasoning. Basically, Hilbert wanted to formulate a formal axiomatic system that would encompass all mathematics. Hilbert hoped that this was the way to obtain the greatest possible objectivity and exactness in mathematics. He says, "There can no longer be any doubt about proofs. The deductive method should be completely clear" [2]. Robert Rosen has a condensed comment about this attitude [3]: "Hilbert and his formalistic school argued that what we have called *semantic truth could always be effectively replace by more syntactic rules*. In other words, any external referent, and any quality thereof, could be pulled into a polysynaptic system. By a purely syntactic system, they understood: (1) a finite set of *meaningfulness* symbols, an alphabet; (2) a finite set of rules for combining these symbols into strings or formulas; (3) a finite set of production rules for turning given formulas into new ones. In such a purely syntactic system, *consistency is guaranteed*" (italics added). In 1931 Gödel and his incompleteness theorems effectively demolished the formalist program.

Kurt Friedrich Gödel was born in Brno (Czech Republic). Gödel's childhood was generally a happy one though he was timid and could get upset easily. When he was six- or seven-years old, Kurt had rheumatic fever, and despite eventual full recovery, he started to believe that he had suffered permanent heart damage as well. These were the early signs of Gödel's later preoccupation with his health. From 1916 to 1924, he carried on with his schooling at the Deutches Staats-Realgymnasium where he proved himself an outstanding student and received the highest marks in all subjects. After graduation in 1924, Gödel went to Vienna to begin his studies at the University. Vienna was his home for the next fifteen years, and he also became an Austrian citizen in 1929. At the University, Gödel was indecisive about whether he should study mathematics or physics at first. It is said that Gödel's decision to concentrate on mathematics was because of his taste for precision and a great impression that one of his professors, the number-theorist Philipp Furtwängler, made on him. A description of the mathematical scene at the University of Vienna in those days is given by Olga Taussky-Todd in her reminiscences of Gödel [4]— Gödel hardly ever spoke but was very quick to see problems and to point the way to a solution. It was evident that he was exceptionally talented. In her memoires, Taussky said that one could talk to Gödel about any problem; he was always very clear about what was the issue and explained matters slowly and calmly. Gödel's principal teacher was Hans Hahn, a mathematician of the new generation, who was interested in logic, foundations of mathematics, and philosophy of science. He is best known for the Hahn-Banach extension theorem.

Those were the days of the beginnings of what later became known as *The Vienna Circle* (Der Wienner Kreis). Admission to its meetings was by invitation only, and it was Hahn who invited Gödel to the sessions of the circle. Gödel first appeared in 1926 when the circle was engaged in the second reading of Wittgenstein's *Tractatus* [5]. From then until 1928, he attended meetings regularly and only occasionally thereafter. In later years, he had to admit that, from the very beginning of his participation, he was not in sympathy with the circle's views. Gödel maintained regular contact with some of the members of the Vienna Circle after 1928, especially with Rudolf Carnap. He turned his back on the Viennese empiricism of Hahn and Moritz Schlick with deep conviction that there were truths to be found not just in the empirically perceivable, but perhaps beautiful and enduring in the realm of abstract conceptions, where they awaited human discovery not through tangible perception but by thought alone. It seemed that Gödel's direction of creative work was mostly influenced by Carnap's lectures on mathematical logic and the book *Grundzüge der theoretischen logick* (Principles of theoretical logic) [6] by David Hilbert and Wilhelm Ackermann. An open problem at the time was the question whether a certain system of axioms for the first-order predicate calculus was complete—that is, whether every logically valid statement was a theorem of the predicate calculus. Gödel came up with a positive solution to the completeness problem, and that significant achievement commenced his research career. His research was completed in the form of his doctoral dissertation at the University of Vienna in the summer of 1929. The degree was granted in February 1930. This year was Gödel's *annus mirabilis* when he, for the first time, announced his great result publicly. It was during the conference on the foundations of mathematics in Königsberg in 1930 when Gödel, a relatively unknown graduate student, said a few words indicating that he had a proof for incompleteness of arithmetic. Basically, he was ignored by everyone present except for one mathematician whose name was John von Neumann. It happened at one of the final sessions. The rest is history.

The scientific community accepted his results steadily but slowly. Still, young Gödel had to spend time in psychiatric hospitals. He had a history of nervous breakdowns caused by hard work. In the war year of 1940, Kurt Gödel finally arrived in the United States after he had traveled over Siberia, Japan, and vast Pacific. He went to Princeton and never returned to Vienna. He remained bitter about his status in Austria from 1939 to 1940 and placed the blame for this situation more on Austrian "schlamperei" than on the outrageous Nazi conditions. On his $60^{th}$ birthday in 1966, he turned down an honorary membership in the Austrian Academy of Sciences.

In summary, there were two mathematical systems in the first third of the twentieth century. They were so extensive that it was generally assumed that every mathematical proposition could be either proved or disproved within the system. The great logician Kurt Gödel amazed the entire mathematical world with the paper [7] published in 1931 showing that it was not the case that every mathematical proposition could be either proved or disproved within these systems. The paper begins with the following words:

> The development of mathematics in the direction of greater precision has led to large areas of it being formalized, so that proofs can be carried out according to a few mechanical rules. The most comprehensive formal systems to date are, on the one hand, the *Principia Mathematica* of Whitehead and Russell, and, on the other hand, the Zermelo-Fraenkel *System of axiomatic set theory*. Both systems are so extensive that all methods of proof used in mathematics today can be formalized in them, i.e., can be reduced to axioms and rules of inference. It would seem reasonable, therefore, to surmise that these axioms and rules of inference is sufficient to decide all mathematical questions that can be formulated in the systems concerned. In what follows it will be shown that this is not the case, but rather that in both of cited systems, there exist relatively simple problems of the theory of ordinary whole numbers which cannot be decided based on the axioms [7].

Gödel explained that the theorems he would prove did not depend on the special nature of two systems under consideration but rather held for an extensive class of mathematical systems. Gödel's results belong to logic, mathematics, philosophy, and physics. His work has made important contributions to proof theory connecting classical logic, intuitionistic logic, and modal logic and to constructive mathematics and the set theory.

## 2.2. An informal proof of Gödel's incompleteness theorems of formal arithmetic

The understanding of a proof requires certain knowledge of mathematical logic, so we assume that the reader is familiar with basic concepts. Kurt Gödel published his proof of the theorem of incompleteness at the University of Vienna in 1931.

Let $K$ be the first-order predicate theory. What Gödel proved in his Ph.D. thesis is the *theorem of completeness*. A formula $A$ is valid if and only if $A$ is a theorem of $K$. In what follows, the set of natural numbers is denoted by $N = \{0,1,\ldots,n,\ldots\}$; thus, we have an algebra $N = (N, +, \cdot, 0, 1)$ that is the *standard model* of the *formal theory of numbers* (*formal arithmetic*) N.

Formal theory of numbers is a first-order theory with equality that has one relation symbol $R_1^2$, a constant $a_1$, and function symbols $f_1^1, f_1^2, f_1^3$. Instead of $R_1^2, a_1, f_1^1(t_1), f_1^2(t_1, t_2), f_1^3(t_1, t_2)$, we write $=, 0, t_1', t_1 + t_2, t_1 \cdot t_2$. Proper axioms are: (1) $(x = y \wedge y = z) \Rightarrow x = z$, (2) $x = y \Rightarrow x' = y'$, (3) $x' \neq 0 (\neg x' = 0)$, (4) $x + 0 = x$, (5) $x + y' = (x + y)'$, (6) $x \cdot 0 = 0$, (7) $x \cdot y' =$ $x + xy$, (8) $(A(0) \land (\forall x)(A(x) \Rightarrow A(x'))) \Rightarrow (\forall x)A(x)$, where $A(x)$ is a formula of N. As it is already known, (8) is called the *axiom of induction*. The terms $0, 0', 0'', 0''', \ldots$ are called *numerals* and denoted by $0, [1], [2], [3], \ldots$ In general, if $n$ is a nonnegative integer, we will let $[n]$ stand for the corresponding numeral $0''^{\cdots'}$, i.e., for 0 followed by $n$ strokes. Using above axioms, one can, for example, prove that in N $\vdash [2] + [2] = [4]$.

A number-theoretic relation $R(x_1, \ldots, x_n)$ is said to be *expressible* in N by a formula $R(x_1, \ldots, x_n)$ of N with $n$ free variables such that, for any natural numbers $k_1, \ldots, k_n$,

1. if $R(k_1, \ldots, k_n)$ is true, then $\vdash_N R([k_1], \ldots, [k_n])$;

2. if $R(k_1, \ldots, k_n)$ is false, then $\vdash_N \neg R([k_1], \ldots, [k_n])$.

For a number-theoretic function $f(x_1, \ldots, x_n)$, we say that it is *representable* in N if and only if there is a formula $F(x_1, \ldots, x_n, x_{n+1})$ of N with the free variables $x_1, \ldots, x_n, x_{n+1}$ such that, for any $k_1, \ldots, k_{n+1}$,

if $f(k_1, \ldots, k_n) = k_{n+1}$, then $\vdash_N F([k_1], \ldots, [k_n], [k_{n+1}])$

$\vdash_N (\exists_1 x_{n+1}) F([k_1], \ldots, [k_n], x_{n+1})$, where $\exists_1 x_{n+1}$ means "there is only one $x_{n+1}$."

If $R(x_1, \ldots, x_n)$ is a relation, then the *characteristic function of R*, $c_R(x_1, \ldots, x_n)$ is defined as

$$c_R(x_1, \ldots, x_n) = \begin{cases} 0 & \text{if } R(x_1, \ldots, x_n) \text{ is true} \\ 1 & \text{if } R(x_1, \ldots, x_n) \text{ is false} \end{cases}. \qquad (2.2.1)$$

It is easy to see that $R(x_1,\ldots,x_n)$ is expressible in N if and only if $C_R(x_1,\ldots,x_n)$ is representable in N. Namely, if $R(x_1,\ldots,x_n)$ is expressible in N by a formula $F(x_1,\ldots,x_n)$, then it is easy to verify that $C_R(x_1,\ldots,x_n)$ is representable in N by the formula $(F(x_1,\ldots,x_n) \wedge x_{n+1} = 0) \vee (\neg F(x_1,\ldots,x_n) \wedge x_{n+1} = [1])$. Conversely, if $C_R(x_1,\ldots,x_n)$ is representable in N by a formula $B(x_1,\ldots,x_n,x_{n+1})$, then $R(x_1,\ldots,x_n)$ is expressible by the formula $B(x_1,\ldots,x_n,0)$.

Representability of functions and relations in N leads to a class of number-theoretic functions that are of great importance in mathematical logic. The following function mappings $N^k \to N$, for all $k \in N$, are called *initial functions*: (1) the zero function $Z(x) = 0$, for all $x$; (2) the successor function $x' = x + 1$, for all $x$; (3) the projection function $I_i^n(x_1,\ldots,x_n) = x_i$, for all $x_1,\ldots,x_n$.

The following are rules for obtaining new functions from given ones:

1. Substitution: $f(x_1,\ldots,x_n) = g\bigl(h_1(x_1,\ldots,x_n),\ldots,h_m(x_1,\ldots,x_n)\bigr)$, $f$ is said to be obtained by *substitution* from the functions $g(y_1,\ldots,y_n), h_1(x_1,\ldots,x_n),\ldots, h_m(x_1,\ldots,x_n)$.

2. Recursion: $f(x_1,\ldots,x_n,0) = g(x_1,\ldots,x_n); f(x_1,\ldots,x_n,y_{n+1}) = h\bigl(x_1,\ldots,x_n,y,f(x_1,\ldots,x_n,y)\bigr)$. If $n = 0$, we have $f(0) = k, f(y+1) = h\bigl(y,f(y)\bigr)$; $f$ is said to be obtained from $g$ and $h$ by *recursion*.

3. $\mu$ −operator: Assume that $g(x_1,\ldots,x_n,y)$ is a function such that for any $x_1,\ldots,x_n$ there is at least one $y$ such that $g(x_1,\ldots,x_n,y) = 0$. We denote the least number $y$ such that $g(x_1,\ldots,x_n,y) = 0$ by $\mu_y(g(x_1,\ldots,x_n,y) = 0)$. Let $f(x_1,\ldots,x_n) =$ $\mu_y(g(x_1, \ldots, x_n, y) = 0)$. Then $f$ is said to be obtained from $g$ by means of the $\mu$ −operator if the given assumption about $g$ holds.

A function $f$ is said to be *primitive recursive* if and only if it can be obtained from the initial functions by any finite number of substitutions and recursions. A function $f$ is said to be *recursive* if and only if it can be obtained from the initial functions by any finite number of applications of substitutions, recursions, and $\mu$ −operator. Almost all elementary number-theoretic functions we know are primitive recursive. However, there are recursive functions that are not primitive recursive.

A relation $R(x_1, \ldots, x_n) \subseteq N^n$ is *recursive* if $c_R(x_1, \ldots, x_n)$ is a recursive function. It is very important for the proof of Gödel's theorem to have the following assertion: *every recursive relation is expressible in* N. This is a corollary of the theorem: *every recursive function is representable in* N.

Now we are ready to introduce the ingenious idea of Gödel, the so-called *Gödel numbering*. Let T be a first-order theory. With each symbol $v$ of T, we correlate a positive integer $g(v)$ called Gödel number of $v$ in the following way:

$$g\big((() = 3, g())\big) = 5, g(,) = 7, g(\neg) = 9, g(\Rightarrow) = 11;$$

$$g(x_k) = 5 + 8k, \quad \text{for } k = 1, 2, \ldots;$$

$$g(a_k) = 7 + 8k, \quad \text{for } k = 1, 2, \ldots; \quad (2.2.2)$$

$$g(f_k^n) = 9 + 8(2^n 3^k), \quad \text{for } k, n \geq 1;$$

$$g(A_k^n) = 11 + 8(2^n 3^k), \quad \text{for } k, n \geq 1.$$

Obviously, different symbols have different Gödel numbers, and every Gödel number is an odd positive integer. *Examples.* $g(x_2) = 21$, $g(a_4) = 39$, $g(f_1^2) = 105$, $g(A_2^1) = 155$. If $u_1 u_2 \ldots u_r$ is an expression, we define Gödel number to be $g(u_1 u_2 \ldots u_r) = 2^{g(u_1)} \cdot 3^{g(u_2)} \ldots p_{r-1}^{g(u_r)}$, where $p_i$ is the $i^{th}$ prime and $p_0 = 2$. For example, $g(A_1^2(x_1, x_2)) = 2^{g(A_1^2)} \cdot 3^{g(()} \cdot 5^{g(x_1)} \cdot 7^{g(,)} \cdot 11^{g(x_2)} \cdot 13^{g())} = 2^{107} \cdot 3^3 \cdot 5^{13} \cdot 7^7 \cdot 11^{21} \cdot 13^5$.

Different expressions have different Gödel numbers, by the uniqueness of the factorization of integers into primes. If we have an arbitrary finite sequence of expressions $u_1, u_2, \ldots, u_r$, we can assign a Gödel number to this sequence by setting $g(u_1, u_2, \ldots, u_r) = 2^{g(u_1)} \cdot 3^{g(u_2)} \ldots p_{r-1}^{g(u_r)}$.

Let T be any theory with the same symbols as N. Then T is said to be $\omega$-consistent if and only if, for every formula $A(x)$ of T, if $\vdash_T A([n])$ for every natural number $n$, then it is not the case that $\vdash_T (\exists x) \neg A(x)$.

It is easy to see that if T is $\omega$-consistent, then T is consistent. Namely, if T is $\omega$-consistent, consider any formula $A(x)$ that is provable in T, e.g., $x = x \Rightarrow x = x$. In particular, $\vdash [n] = [n] \Rightarrow [n] = [n]$, for all natural numbers $n$. Therefore, $(\exists x) \neg (x = x \Rightarrow x = x)$ is not provable in T. T is consistent by the tautology $\neg A \Rightarrow (A \Rightarrow B)$ because in case T being inconsistent, every formula $B$ would be provable in T.

Define a relation $R := \text{Proof} \in N \times N$ in the following sense: $\text{Proof}(a, b) \leftrightarrow b$ is the Gödel number of a proof of the formula whose Gödel number is $a$. $\text{Proof}(x_1, x_2)$ is recursive relation because of the uniqueness of Gödel numbering. Let $Proof(x_1, x_2)$ be a formula that represents the relation $\text{Proof}(x_1, x_2)$ in N. $Proof(x_1, x_2)$ is a formula $A_1^2(x_1, x_2)$ of N, of course. The following formula of N

$$(\forall x_2)\neg Proof(x_1, x_2) \tag{2.2.3}$$

is crucial in these considerations. Let $m$ be the Gödel number of the formula (2.2.3) and let

$$(\forall x_2)\neg Proof(\lceil m \rceil, x_2) \tag{2.2.4}$$

be a formula in the language of N.

*Theorem of Incompleteness* [7].

(1)          If N is consistent, then $\nvdash (\forall x_2)\neg Proof(\lceil m \rceil, x_2)$.

(2)          If N is $\omega$-consistent, then    $\nvdash \neg(\forall x_2)\neg Proof(\lceil m \rceil, x_2)$.

A proof could be demonstrated in the following way:

(1) Suppose N is consistent and $\vdash (\forall x_2)\neg Proof(\lceil m \rceil, x_2)$. Let $k$ be the Gödel number of a proof of this formula. Therefore, $Proof(m, k)$ is true and thus

$$\vdash Proof(\lceil m \rceil, \lceil k \rceil).$$

However, from the assumption, we have $\vdash (\forall x_2)\neg Proof(\lceil m \rceil, x_2)$. From the logical axioms of N, we have, for $x_2 = \lceil k \rceil$, $\vdash \neg Proof(\lceil m \rceil, \lceil k \rceil)$. N is consistent, thus a contradiction.

(2)      Assume that N $\omega$-consistent and that $\vdash \neg(\forall x_2)\neg Proof(\lceil m \rceil, x_2)$ i.e.,

$\vdash (\exists x_2) Proof(\lceil m \rceil, x_2)$. N is also consistent, so that not $\vdash (\forall x_2)\neg Proof(\lceil m \rceil, x_2)$. Therefore, for every natural number $n$, $n$ is not the Gödel number of a proof in N of $(\forall x_2)\neg Proof(\lceil m \rceil, x_2)$, i.e., for every $n$, Proof$(m,n)$ is false. So, for every $n$, $\vdash \neg Proof([m],[n])$. Let $A(x_2) := \neg Proof([m], x_2)$. Then for every $n$, $\vdash A([n])$. By $\omega$-consistency of N, it follows that not

$\vdash (\exists x_2)\neg\neg Proof([m], x_2)$. Hence, not $\vdash (\exists x_2)Proof([m], x_2)$. But this is a contradiction with

$\vdash (\exists x_2)Proof([m], x_2)$.

As a consequence of the Theorem of incompleteness, Gödel proved the following:

*Theorem of consistency of* N [8]. If N is consistent theory, there is no proof in N of its consistency.

This fact ruined Hilbert's Program. To conclude, if we can prove that $\boldsymbol{N} = (N, +, \cdot, 0, 1)$ is a model of N, then we have consistency of N. Therefore, we "believe" that $\boldsymbol{N}$ is the standard model of N.

## 2.3. Gödel's incompleteness theorems as a metaphor. Real possibilities and misrepresentation in their application

Gödel's theory states that there are the limitations of mathematics that does not mean that it is flawed in any way. More generally, it tells us about the limitations of our knowledge or limitations of broadening or completing our knowledge. Gödel's theorem is an example of vertical progress in mathematics, just as Einstein's theory of relativity and Planck's separation of the discrete from the continuous world are examples in physics (see subchapter 1.2). Their discoveries can be described as an improvement of something that already existed to something new (but very surprising). What these contributions have in common is that they have moved the boundaries of science and mathematics. Metaphorically speaking, as in the novels of magical realism, suppose the spirit were to appear during your dinner. Not surprisingly, these circumstances would look unrealistic to you. But if you said "Spirit, this cake is delicious," it would already be magical realism—that is, you would accept the new situation. If you understood and adapted to this newly emerged world, your perception would be exceptional (i.e., you would have the ability to make vertical progress) (see Pekić's comment in subchapter 1.1). So, there is no mixing of different levels of understanding, just a movement towards a higher level while these steps are strongly connected by the principle of correspondence. In other words, in the region as far as physics has come, the question is not whether there are "holes" but where the limitations of knowledge in physics are.

After his discovery in 1931, Gödel was misrepresented by postmodernists, particularly by philosophers who wanted to show that everything was meaningless. Gödel's reasoning and intuition opened the door for "unproved truths" what seemed to be extremely counterintuitive. Mathematicians and philosophers did not recognize the deep consequences of Gödel's results at first. As time went on, experts in other sciences noticed similar situations in their own fields, and Gödel's theorem became *a metaphor* for the systems whose consistency could not be proved within themselves. What did Gödel invent in the abstract universe of mathematics? He proved that 1. there exist matters in that universe that cannot be proved nor disproved, such as axioms; 2. there is no possible way to prove these matters *within* that universe. Basically, this is a problem of self-reference (an issue seen in Russell's paradox about sets) expressed through the famous self-referential sentence (liar paradox): "This sentence is false" creating a logical circularity. If the sentence is true, then it cannot be true; if it is false, it must be true. Gödel applied a similar logic to the whole system of mathematics but with the sentence "This statement is unprovable" and then made a conversion into a number statement about numbers by using a code system later entitled "Gödel numbering." Finally, he found that this proposition could not be proved within system. In the previous subchapter, we gave an informal proof of these theorems. For further reading and a specific look at the insight of Gödel's theorems, we recommend papers by Hu [9] and Johnstone [10].

Since the lair paradox is an apparently trivial problem, the question is why it is a deep problem. Because its solving is a part of a broad project of understanding truth that involves finding a theory of truth, definition of truth, or a proper analysis of the concept of truth [11]. Therefore, although we have no evidence about direct applications of Gödel's theorems in physics, they are deeply rooted indirectly. Perhaps the following reasoning has the attributes of a paradox. Despite the fact that Gödel's incompleteness theorems have not yielded results in physics so far, they are perhaps much more important for physics than for mathematics itself. They refer only to a part of mathematics, however, in physics these theorems, despite their inapplicability, can contribute to defining the boundaries of knowledge. In his famous book *Gödel, Escher, Bach: An Eternal Golden Braid* [12], Douglas Hofstadter illustrates Gödel's world by following his own preference for using metaphors, analogies, images, and so forth [13].

Let us now return to subchapter 1.1 where we mention some examples of how one can question the possible existence of the theory of everything. We also mention Gödel's work in the context of limits supporting "the end of physics" in subchapter 1.2. One should always have in mind that Gödel's theorem deals with formal systems, and it is questionable whether physical theories are such systems in the mathematical sense. One way to implement Gödel's work is to map the quantum field theory onto number theory [14]–[15], which is possible under certain assumptions. More often, physicists look for an inspiration in Gödel's work and work in *analogy* although sometimes they do not point out that it is just an analogy. Murphy [16] gives several examples of physicists who work on "incompleteness" of physics that has nothing to do with Gödel's incompleteness. However, in other cases they admit that it is just an analogy that might lead to new results [17]. One can say that Gödel's theorems can be treated as a guideline and, as such, are very valuable.

**2.4. Gödel's excursions into physics problems and computer science**

Physicists, science philosophers, and historians all agree that Einstein's theory of relativity (special and general) is one of the greatest achievements of the human mind not only in the twentieth century but of all time (see subchapter 1.3). Kurt Gödel was always interested in physics and established a very good relationship with Albert Einstein in Princeton. They had long conversations about relativity theory, so it came as no surprise when Gödel offered his contribution to general relativity. Many compliments are given to it because it is mathematically consistent but still leads to many interesting consequences. Actually, his contribution to physics just includes a scientific paper published in the most valuable journal *Reviews of Modern Physics* and more philosophically inclined work in a collection of papers [18]–[19].

In order to explain his contribution and its importance shortly, we must introduce some concepts and accompanying terminology. We use the term *reference frame* for a system of coordinates within which we determine the position of an object. Usually, it is the standard Cartesian system with coordinates *x, y, z* or, more conveniently, $x_1, x_2, x_3$. If we add time *t* to the coordinates, this combination defines *an event.* Theory of special relativity is based on two postulates that summarize results from experiments. One is that the laws of physics appear no different in all reference frames moving uniformly (with constant speed) with respect to each other.

The other postulate is that the velocity of light in the vacuum is always the same independently of the possible motion of the light source or the observer, which is the largest possible velocity in the universe. If one respects these two postulates, one can find the most important rules for the transformation of physical quantities when the observer moves from one reference frame to another. Usually, we consider two reference frames with parallel axes and assume that one of the frames ($S'$) moves with respect to the other one ($S$) with constant (uniform) velocity along one of the axes (e.g., $x_1$). We need to determine how the coordinates and time in the system $S'$ are expressed in terms of the coordinates and time in $S$. In classical physics, these are so-called Galilean transformations, but they take a different form here: Lorentz transformations. $x_1'$ is expressed in terms of $x_1$ and $t$, but more importantly, we have a new quality: $t'$ that is expressed in terms of both $x_1$ and $t$. This implies that time flows differently in a moving frame, which is one of the many fascinating consequences of Lorentz transformations.

One of the most important ideas of relativity is that spatial coordinates and time are completely equivalent in the description of any object or phenomenon, which means that the time should be treated as the fourth coordinate together with three spatial coordinates. (Actually, to have the same dimensionality, the fourth coordinate is *ct.*) So, we deal with so-called *space-time* instead of just space. It is the three-dimensional Euclidean space with fourth coordinate added. In such a four-dimensional generalization (Minkowski space), each point is defined by these four coordinates. The set of these points describing certain motion defines a *worldline.* We choose the set of straight lines for which the distance from the origin equals precisely *ct*, and this set defines the border of a *light cone.* This is illustrated in figure 2.1 for a space with two spatial dimensions and light. The worldlines lying inside the cone are entitled time-like, and those outside the cone are called space-like.

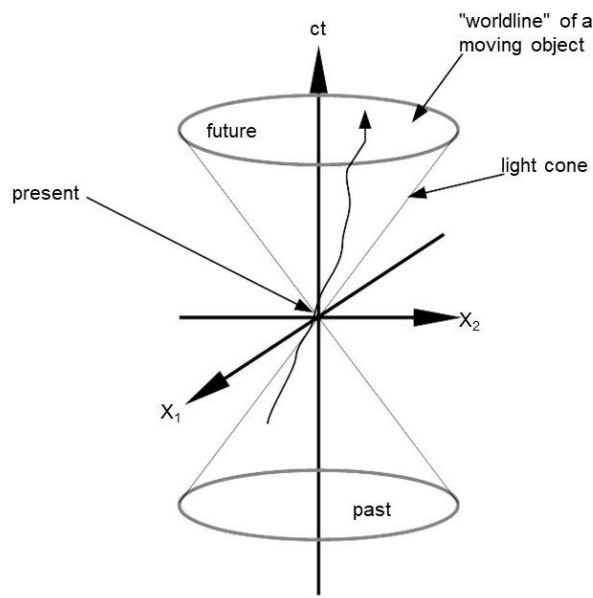

Figure 2.1. Minkowski space with two spatial and one temporal coordinate interval. (Figure courtesy of Miloš Kapor.)

Albert Einstein generalized the special relativity to the general theory of relativity by including gravity in the following way: since the origins of gravitational force are masses, he assumed that the presence of masses changes the geometry of the space-time. Mathematically, he related a *metric tensor* describing the shape of the space-time (its curvature) to the tensor describing the distribution of mass and energy in this space by using Einstein's field equations. There are not many exact solutions of these equations. Among them, some describe the expanding universe that agrees well with present experimental data (i.e.*,* astronomical observations).

Now we come to Gödel's contribution. Gödel used Einstein's equations to propose a model of the rotating universe for which there also exists an exact solution. By itself, it is an important achievement, however, there is more. This solution has one very strange property: it allows the existence of time-like worldlines that are closed. There are many physical arguments for the existence of an anisotropy in time implying that time flows only in one direction that we choose to call the future (see also subchapter 3.1). It is conveniently named the arrow of time. The solutions proposed by Gödel indicate that motion can occur in both directions in time, or better formulated, that there is no privileged direction in time. For the moment, this is treated as a speculation since this solution does not lead to the expanding universe. Therefore, there is a lot of room for the discussion and various interpretations, and more about this solution and related philosophical questions is discussed in chapter 3.

Just as Gödel's work offered a new insight into relativity, so did his remarks about computer science. John von Neuman was very ill in Spring 1956 (he never recovered and died in 1957), and Gödel tried to distract him by sending him a letter [20] with a question that later became known as "$P = NP$ problem" in computer science. We consider the problems (questions) with answers of "yes – no" type. It is favorable that computing time for solving a problem polynomially depends on the input of the algorithm. This time is called *polynomial time*.

The general class of questions for which an algorithm can provide an answer in polynomial time is denoted by "*P*" or "class *P*." Solutions to these problems can be computed in a polynomial amount of time compared with the complexity of the problem. If the algorithm may proceed according to several possibilities at any step of computation, then such an algorithm is called a *nondeterministic algorithm*. The computation of the nondeterministic algorithm is a tree whose branches correspond to different possibilities. If a particular branch of computation leads to the final state, the algorithm accepts the input. The class of questions for which the nondeterministic algorithm can provide an answer is denoted by $NP$. $P = NP$ means that for every problem in $NP$, there must be an algorithm of polynomial complexity that provides a solution of the problem. There are problems that are more complex than the problems of the class $NP$. One can, by his experience, probability, intuition, or visual thinking, guesses the solution of the extremely complex problem. That was the reason why Gödel considered that humans differed from machines. It might be difficult to find the solutions of $NP$ problems. Computation may take billions of years, but once the solution is found, it is easily checked. The $P = NP$ conjecture is the most important problem in computer science and one of the most significant ones in mathematics. This is one of the *Millennium Prize Problems* selected by the Clay Mathematical Institute that is offering a $1 million prize for the correct solution of any of the problems. The class $NP$ is very rich. There are thousands of $NP$ problems in mathematics, optimization, artificial intelligence, biology, physics, economics, industry, etc. An example of $NP$-class problem is the so-called SAT problem. SAT is the set of satisfiable formulas of sentential logic. The truth-table method for determining whether a formula with n variables is satisfiable involves forming $2^n$ lines of the formula's truth table and examining if there is a line making the formula true. If one microsecond is needed for a line of the truth table of 80 propositional variables, then the execution time of the algorithm is $2^{80}$ microseconds, which exceeds the age of the universe. If $P = NP$, then there is a simple algorithm to decide any formula from SAT. The set of Hamiltonian graphs is in $NP$—that is, we might not know the fast polynomial algorithm, but we do know how to show that the graph has a Hamiltonian cycle. After many decades of the struggle with $P = NP$, we are aware that it is a very difficult problem and that we very much need scientists of Gödel's and von Neumann's caliber to seriously attack this problem. So, Kurt Gödel was far ahead of his time when he posed this problem.

# Chapter 3

## Time in physics

### 3.1. Time in philosophy and physics. Beyond Gödel's time

*Time* is one of the most enigmatic phenomena that we experience and try to explain throughout our lives. It is a *concept*, and only philosophy deals with pure concepts, which is its exclusivity. In his book *Time* [1], Rüdiger Safranski, a German philosopher, presents the understanding of time and its history in a simple manner although time is perhaps the most difficult philosophical problem. Considering different approaches to the concept of time by various philosophers, he offers several perspectives of time: (1) psychological as the most attractive but the most superficial, (2) literary as the closest to philosophy, and (3) natural-scientific as "too short." Many aspects of the above perspectives (time in philosophy, time in physics, functional time, endo-time, etc.) are discussed in Mihailović *et al*. [2]. Isaac Newton introduced what provided the theoretical basis for the Newtonian mechanics: the concept of absolute time and space [3]. According to him, absolute time and space are the independent parts of objective reality: "Absolute, true and mathematical time, of itself, and from its own nature flows evenly regardless of anything external, remains always similar and immovable" [3]. We cite an interpretive text about Kant's views of time [4] offered by Dorato *et al*. [5]: "Kant's views of time, in such a way as to distillate *three conditions* that are *together necessary* for the ideality of time in Kant's sense: (i) time must be *no substantial*, and the resulting relationism must be constructed in such a way that both (ii) the difference between past and future and that (iii) between earlier and later than, must be *mind-dependent*." Schopenhauer was so fascinated by Kant's understanding of time that he noticed that we had "carried" our heads in time and space before Kant, while we "carried" time and space in our heads after Kant. It is also inevitable to mention the work of McTaggart "The Unreality of Time" [6] published in 1908. He is famous for arguing that time is unreal because our descriptions of time are contradictory, circular, or insufficient. Einstein's interpretation of Lorentz's transformation introduced a concept different from absolute time: it was the time that depends on a reference frame.

We use the following points to introduce an important concept about time metaphorically. Because of the reliance on hypotheses, induction, and their limitations, physicists always have the problem of 1. choosing the right direction in research; 2. "dealing" with time. This situation can be visualized on the bifurcation map of the differential logistic equation (the logistic equation is not chosen randomly since it appears to be a type of *master* equation in physics and science [2]). The map consists of black and white spots that correspond to chaos and stability, respectively. Physicists are placed in one of the white spots and trying to move to other white spots not in an arbitrary direction but in the direction that must be defined by the universe, its content, and the way it evolved. This universal direction was called *the arrow of time* (i.e., the "one-way direction" or "asymmetry" of time).

> Let us draw an arrow arbitrarily. If as we follow the arrow we find more and more of the random element in the state of the world, then the arrow is pointing towards the future; if the random element decreases the arrow points towards the past. That is the only distinction known to physics. This follows at once if our fundamental contention is admitted that the introduction of randomness is the only thing which cannot be undone. I shall use the phrase "time's arrow" to express this one-way property of time which has no analogue in space [7].

The arrow of time is a different concept from time itself. It assumes that time has a direction and that there are different arrows of time: 1. *The thermodynamic arrow of time.* This arrow is related to the entropy of an isolated system that either remains constant (a rare case) or increases with time (the second law of thermodynamics) (fig 3.1). Because its direction and entropy are connected, the past (lower entropy) and the future (higher entropy) can be distinguished.

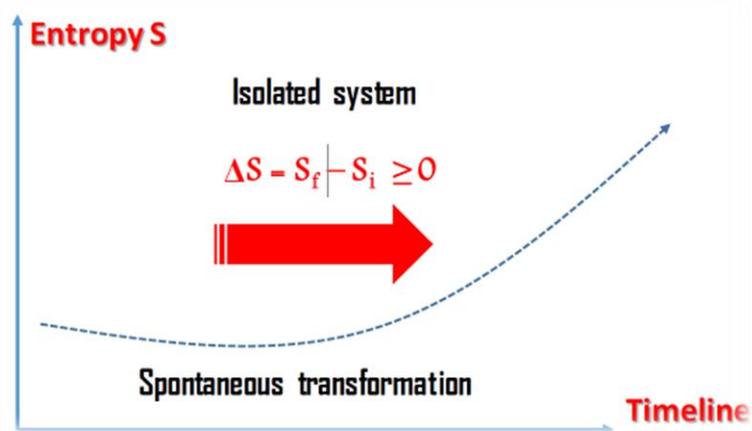

Figure 3.1. Schematic representation of entropy as an arrow of time. (Reproduced by permission from [8].)

2. *Cosmological arrow of time*. Starting from the fact that the arrow of time is defined by the direction of time in which entropy increases, Hawking [9] shows that the cosmological arrow of time is determined by the direction of the time of the universe expansion. The cosmological arrow of time may be associated with the thermodynamic one or even depend on it. 3. *Quantum arrow of time*. Schrödinger's equation that describes the evolution of the wave function (the basic source of information in quantum mechanics) is a linear differential equation of time, so it is irreversible. The same situation holds for the so-called "collapse of wave function" owing to the measurement process that "sends" a system into one particular state. Because of its collapse, the complete information from the wave function is lost without any possibility to be reconstructed. In this regard, the process is time-irreversible, and an arrow of time is formed. It is not clear how the quantum arrow of time is related to others, but it is possibly related to the thermodynamic arrow of time since wave functions in nature show tendency to collapse into higher entropy states, as opposed to lower entropy ones. In quantum physics, the results of measurements in the future are partially constrained by the results of measurements in the past. Accordingly, the concept of a quantum state is unaffectedly "time-asymmetric"; to avoid this concept, asymmetry is separated by "symmetrization of time" in Aharonov *et al.* [10]. They introduced the two-state vector formalism of quantum mechanics and later extended their work [11]–[12]. 4. *Radiative arrow of time*. In theory, equations allow waves to be convergent, whereas this situation has never been observed in nature. This discrepancy in the flow of radiation is regarded as the radiative arrow of time by some scientists. Since radiation implies increased entropy, and convergence implies decreased entropy, this arrow of time may be linked to the thermodynamic arrow of time (consider radiation through its particles in quantum mechanics). 5. *Weak arrow of time*. In particle physics, the CPT (charge conjugation, spatial inversion, and time reversal) theorem asserts that the laws of physics are invariant under the combined C, P, T transformations of any interaction. Empirical observations revealed that they are partially invariant in specific subtle interactions—that is, they are not invariant to CP operations [13]. Consequently, the laws of physics are not completely invariant to time reversal. This indicated another arrow of time named the weak arrow of time. 6. *Psychological arrow of time*. This arrow corresponds to the human perception of time. It was first mentioned by Stephen Hawking in his book *A Brief History of Time* [14] and later in the series of his lectures [15]–[16] delivered at different American universities. Let us begin with the quote from his book [14]: "Disorder increases with time because we measure time in the direction in which disorder increases." This implies that the psychological arrow of time is determined by the thermodynamic arrow of time. Therefore, we can define the direction of the psychological arrow of time in the following way: we *remember* the past (decreased entropy) but *not* the future (increased entropy). This is how we *feel and perceive* the past and the future. This arrow should align with the thermodynamic arrow of time that is well defined [17].

*Beyond Gödel's time.* The model of the universe known as Gödel's space-time [18] (see subchapter 2.4) is based on an essential rotation of matter, but the solution of its equations does not exhibit expansion like the actual universe. Nevertheless, models with both rotation and expansion are also theoretically possible. Precise measurements performed recently indicated that the actual universe does not rotate [19]. This fact neither diminishes nor overshadows questions posed in the context of Gödel's space-time. Gödel's model of the universe, although not the actual model of the universe, is theoretically possible. It allows closed time-like worldlines so that there is no preferred direction of time (see subchapter 2.3). Most philosophers and people from the scientific community did not accept Gödel's view about time based on that solution. In another paper [20], he draws the conclusion that time conjectured by philosophers such as Immanuel Kant is ideal. Richmond [21] interprets this inference as unreal time that does not have objective existence.

Gödel was deeply aware of the ontological nature of time. Then where is Gödel's view about time positioned in philosophy and physics? When we consider Gödel's time, we look at it from the perspective of science or philosophy. In philosophy, the main alternative to presentism, the view that only the present is real, is eternalism according to which all events in the past, the present, and the future already exist. It seems that the most intriguing point is that Gödel's model allows traveling back in time without the exceedance of the local speed of light. While time in physics is defined as irreversible, Gödel's time is cyclic: objects can return at a certain point in the past. To summarize Gödel's time, we use Rigs's response [21] on Kurt Gödel's solution of the Einstein field equations [22]: 1. Gödel's space-time shows that the order of events in time is not always globally consistent in space-times that consist of closed time-like worldlines. 2. Relying on Gödel's model [18], some researchers proposed an idea that time travel was possible without the rotating universe [23]. 3. Our understanding of the laws of physics does not eliminate the possibility of time travel [24]. 4. Since Gödel's space-time results from the Einstein field equations, it is a model of a possible universe [21]. 5. Time travel will turn out to be an open question and be excluded from physics.

Like all open questions in physics, the question of backward time travel may be closed in a manner that is unknown to us at this level of our knowledge. In connection with this dilemma, Borislav Pekić's reasoning [25] about an open question is also interesting: "The existence of spirits in principle does not contradict any law of physics. It is in contrast to the mind that civilization has modeled on empirical evidence for centuries" (see subchapter 1.1).

### 3.2. Does the quantum of time exist?

We dedicated a complete chapter, a lot of additional material in our previous book [2], and subchapter 3.1 to the concept of time, so some other aspects of time in physics are given in this subchapter. Time is treated as a parameter and not as a dynamical variable in quantum mechanics and classical physics. The simplest explanation of why it is not a dynamical variable is that time is not influenced by the action of force in classical physics, special relativity, and quantum mechanics. So, it remains the problem of whether time can be quantized. A popular idea of quantization is that there is the smallest value of a certain quantity ("quantum"), and all other values are multiples of that quantum. This idea is attributed to Max Planck who introduced the quantum of the energy of linear harmonic oscillator to explain the law of the black body radiation. Previously, he used a classical equivalence of the electromagnetic field in a closed volume of space with the system of independent harmonic oscillators; this is also valid for light ("photons") and Millikan's discovery of charge quantization [26].

The obvious question is if there is such a quantity for time. Actually, there is one and it is called the Planck time [27]. First, we introduce the *Planck length* $l_P$

$$l_P = \sqrt{\frac{\hbar G}{c^3}} = 1.616 x 10^{-35} m \tag{3.2.1}$$

that is derived in dimensional analysis from three fundamental physical constants: reduced Planck's constant $\hbar$, gravitational constant $G$, and velocity of light in vacuum $c$. Is this the smallest possible length? The estimated size of a quark is $\sim 10^{-18}$ m; thus, the Planck length is 18 times smaller than the smallest particle in the universe (it is supposed that quarks are composed of smaller particles—preons). Although dimensional analysis is a powerful tool in many situations, it is not a reliable method that can provide us with consistent information in the way that the Planck scale can intrinsically contribute to quantum gravity research with strong relevance [28]. Is this exactly the case with the Planck length? This question is quite natural if one considers the ultimate limits of measurement at this scale [29]. In that case, a refined length requires a large momentum according to Heisenberg's uncertainty principle; thus, its gravitational effect becomes strong. C. Alden Mead [30] finds that it is impossible to measure the position of a particle with the error less than $\Delta x = \sqrt{G} 1.6 x 10^{-35} m$ while analyzing the effect of gravitation on hypothetical experiments. He suggests that it may not be possible to formulate the fundamental length theory that does not include the effect of gravitation in a significant manner. Alden Mead said that Henry Primakoff, David Bohm, and Roger Penrose supported him in the idea that $l_P$ was a fundamental length during his suffering of the referee trouble (1959–1964) [31]. Frank Wilczek later showed that simple dimensional analysis combined with a few elementary facts could lead to profound conclusions [29], [32].

There are various interpretations of this length. The simplest description is that it is the limit where known physical laws may fail since there is no way to go beneath the Planck scale in terms of time or distance, and a new theory combining quantum mechanics and gravity

$$t_P = \frac{l_P}{c} = 5.89 x 10^{-44} s \qquad (3.2.2)$$

is the time taken by light to cross that distance. According to current concepts, this should be the smallest possible time interval that exists between two related events. The Planck time is generally viewed as an elementary "pixel" of time within which the physics of four-dimensional space-time breaks down into much greater number of dimensions assumed by the superstring theory. Our understanding of space-time becomes a little ambiguous beyond this time scale. However, it is not the quantum of time because there is no proof that time is a multiple of the Planck time.

One must remember that the concept of quantization can be introduced in a stricter way: the phenomenon that certain physical quantities have only certain values (the energy of an electron in the hydrogen atom). There is also the spatial quantization of the directions of a particle's angular momentum and spin in the magnetic field. All these appear in nature, and their existence is proved by the numerous studies of the spectra of noninteracting atoms and molecules. We do not know the origin, but we explain their existence by studying the eigenvalue problem of the quantum mechanical operator representing a given physical quantity. The quantization then follows from the requirement that solutions must satisfy certain conditions (mathematical and physical). From this point of view, one needs the quantum of time that, for the moment, should appear only in the theory combining quantum mechanics with gravity [33] (there was popular one such a theory [34] that introduced the quantum of time, so- called chronon).

**3.3. Continuous and discrete time**

The purpose of this subchapter is to summarize how time is treated in dynamical systems in physics and some related problems. The two recognized perspectives of handling time in dynamical systems, finite and infinitesimal, are formalized throughout the two distinct concepts of time—*discrete* and *continuous* time. When it comes to discrete time, the time axis is separated into the smaller time-segments of fixed length, and the resulting number of points and intervals is finite. Quite the opposite, continuous time involves infinitesimally small-time intervals. The changes of a state variable are modeled by the first derivative of the state function, while higher order derivatives are used to formalize higher order changes. Thus, the changes of state variables over time can be related to intervals and specified for specific points or to each point of time. An interesting fact is that continuous time can be treated as an imaginary quantity (*imaginary time*). This time is not imaginary in a literal sense; it is just real time that undergoes the Wick rotation. In the theory of many-body systems, the time dependence of physical quantities in Matsubara's Green functions is described by imaginary time. This allows simpler calculations, and Green's functions are related to the standard ones afterward to evaluate relevant physical properties of a system by using well-elaborated methods (see [35]). This time is a mathematical representation of time that appears in special relativity and quantum mechanics. It may also appear in quantum statistical mechanics and certain cosmological theories.

Physicists need to decide on the treatment of time because the conceptualization of time in dynamical models is not straightforward. The question whether we should use continuous or discrete time includes dilemmas that are far from being resolved (for further reading see [2]). Many symmetries in continuous models are lost after discretization. Small discrete systems are much easier to deal with, while continuum models are more comfortable to use. Regardless of whether nature is essentially discrete or not, most physical models are continuous; therefore, ordinary or partial differential equations are *substituted* with appropriate difference equations. It is more accepted to use discrete difference equation(s) for model building to avoid the process of 1. finding a differential equation to approximate a discrete situation; 2. approximating that differential equation with a difference scheme (for numerical computing purposes) [36], [37]. If some phenomena are already described by equation(s), the corresponding laws can be deduced from symmetry conditions. Since this is not usually the case, we often determine equations from some hypotheses and experimental data [38]. Even if know which equations should be used, we can still be uncertain about the values of parameters because of numerous simplifications that we designed. A mathematically correct solution of the corresponding system of differential or difference equations is not always physically plausible. One reason is that mathematicians and physicists differently understand the term random. The second one is that the solutions of the corresponding system of differential equations may be mathematically correct but physically meaningless [2]. For example, under some conditions that occur in the atmosphere, the energy balance equation can be written in the form of

$$X_{n+1} = A_n X_n - B_n X_n^2, \tag{3.3.1}$$

where $X_n$ is the dimensionless environmental interface temperature, while coefficients $A_n$ and $B_n$ are dimensionless parameters. The solution of (3.3.1) for the specific choice of $A_n$ and $B_n$ is depicted in fig. 3.1 (*upper*) (chaotic behavior). According to Kreinovic's opinion [38], the question is whether we can find domain(s) for this equation (fig. 3.1 *lower*) where physically meaningful solutions exist.

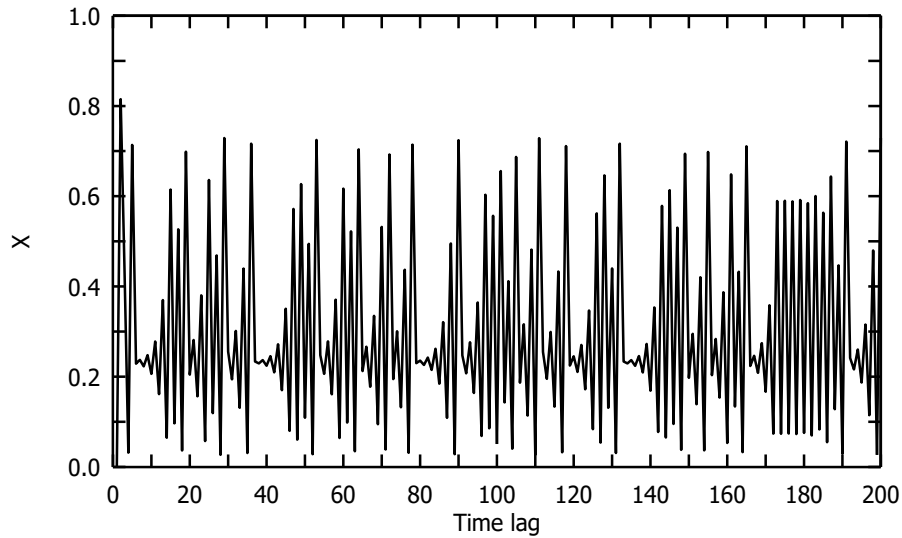

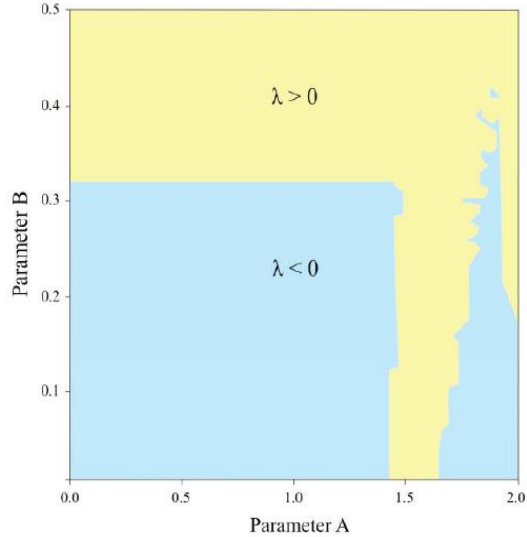

Figure 3.2. Chaotic fluctuations of dimensionless environmental interface temperature ($X$) in (3.3.1) (*upper*). The regions of stable $\lambda < 0$ and unstable $\lambda > 0$ solutions of (3.3.1) determined by the values of Lyapunov exponent ($\lambda$) (see next chapter) and the coefficients $A \in (0, 2)$ and $B \in (0, 0.5)$ (*lower*). (Reproduced by permission from [2].)

There are also constraints on choosing the time step $\Delta t$ because it must satisfy some physical and numerical conditions. In order to solve certain partial differential equations numerically, Courant–Friedrichs–Lewy's (CFL) condition [39] must be met since it is a necessary condition for convergence. According to this criterion, a simulation gives incorrect results if the time step is not less than a certain period of time in many explicit time-marching computer simulations. The choice of the time step depends on how we choose to discretize the system of equations, but spatial and temporal resolutions are connected through the CFL condition.

Surfaces often have such thermal characteristics that the coefficients of the equation (3.3.1) may change significantly during the calculation of the environmental interface temperature from the energy balance equation [2]. All energy transfer processes on the interface occur in the finite time interval. If certain conditions are met, this equation can be transformed into the logistic equation

$$X_{n+1} = rX_n(1 - X_n),$$
$$r = 1 + \tau, \quad (3.3.2)$$
$$\tau = \frac{\Delta t}{\Delta t_p},$$

where $X_n$ is the dimensionless temperature, $0 < r < 4$, $\tau$ is the dimensionless time, $\Delta t$ is the time step, and $\Delta t_p$ is the *scaling time* of energy exchange at the environmental interface. The scaling time can take the form of $\Delta t_p = C_g/\Sigma$, where $C_g$ is the heat capacity of the environmental interface per unit area, while $\Sigma$ is the net energy amount at the environmental interface. $\Delta t_p$ defined in this manner can be used as the criterion for the choice of the time step $\Delta t$ to solve the energy balance equation numerically (i.e., $\Delta t < 2.57\Delta t_p$) [2]. Using this inequality, we avoid the situation when the environmental interface cannot accept an enormous amount of energy that enters a system unexpectedly. Evidently, this criterion depends only on the net energy of the environmental interface and its thermal properties. In the inequality $\Delta t < 2.57\Delta t_p$, the scaling time $\Delta t_p$ is *relaxation time* often accoutered in the theory of nonequilibrium processes. When the time step $\Delta t$ is smaller than the relaxation time, we can observe the energy exchange at a smaller scale (i.e., we can avoid chaos).

### 3.4. Time in complex systems

During the preparation of this book, the Nobel Prize in Physics 2021 was awarded to three scientists "for groundbreaking contributions to our understanding of complex systems" [40]. Syukuro Manabe and Klaus Hasselmann were awarded "for the physical modelling of Earth's climate, quantifying variability and reliably predicting global warming" [40], and Giorgio Parisi was awarded "for the discovery of the interplay of disorder and fluctuations in physical systems from atomic to planetary scales" [40]. This was the first time that the Nobel Prize in Physics was honored to scientists for the research on complex systems, including the climate. In the paper [41] in *Scientific American,* Daniel Garisto says, "Unfortunately, by grouping seemingly unrelated research under the vague umbrella of complex systems, the Nobel Committee for Physics puzzled many observers and led to hazy headlines such as 'Physics Nobel Rewards Work on Climate Change, Other Forces.' What links these very different discoveries is, at first, far from clear. But a close examination reveals some connections—about the aims of science and how scientists can tackle seemingly intractable problems." Reactions to the decision of the Nobel Committee were different, which could have been expected. Among them, one reaction was interesting. "There is no clear definition of complex systems," says Kunihiko Kaneko, a physicist at the University of Tokyo. "But roughly speaking, there are many interacting elements, and they often show chaotic or dynamic behavior" [41]. This view is at least encouraging to think about. The Nobel Committee for Physics explained what had motivated them to make this decision and award the scientists who examined "the phenomena we observe in nature emerge from an underlying disorder, and that embracing the noise and uncertainty is an essential step on the road towards predictability" [41]. If we accept that chaos and disorder are synonyms and recognize philosophical elements that we use to describe connections in this text, then we can more clearly understand Žoze Saramago's credo in the novel *The Doubled* [42]: "The chaos is order yet that to be puzzled out." Our view on the above-mentioned issues and comments can be found in subchapters 1.2 and 1.3. It should be noted that Ilya Prigogine was the first who gave prominence to complex systems and new structures that emerge owing to the internal self-reorganization in his pioneering papers. He was awarded the Nobel Prize in Chemistry 1977 "for his contributions to non-equilibrium thermodynamics, particularly the theory of dissipative structures" [43]. Despite the fact that chemistry was nearby to being closed at the time, chemists noticed the deep nature of his ideas much earlier than physicists.

The above condensed text about people who have built the foundations of complex systems is a short introduction to the *time of complex systems* (or how it is briefly and not quite correctly called *complex time*). It has a fundamental meaning because many questions about the temporal asymptotic behavior of complex systems in infinity cannot be answered by performing finite computations. The broad consensus coming from numerous sources is that the time in complex systems can be described as follows: "Time in complex systems operates concurrently at different scales, runs at multiple rates, and integrates the function of numerous connected systems" [44]. This definition is a framework that includes components extending from obvious to hidden but important ones. Its potential drawbacks are discussed in some papers about complex systems whose number is not negligible. We extend this definition by adding our own experience regarding this quantity at the complex system spatial scales mentioned in the conclusion of the Nobel Committee for Physics in 2021.

Simulation of complex systems is an important branch that allows us to simulate a complex process and integrate a set of underlaying evolution equations. It is difficult for scientists to create an optimal model because interactions between components in a complex system change over time. This becomes even more complicated when the underlying process is unknown. We explain three contexts that reflect the uses of time in complex systems modeling: 1. *Time stepping*. In climate models, the partial differential equations that describe the physical phenomena in the complex earth-atmosphere interaction are discretized numerically in space and time to obtain a solution; therefore, the emerging time-integration strategy (time-stepping with a *time step* $\Delta t$) is essential for model building [45]. This procedure is complex and includes a. different numerical algorithms for the adiabatic fluid dynamics and sub-grid scale diabatic physics; and b. numerical stability limits of each atmospheric process. The time integration in climate models is performed for each $\Delta t$, including smaller time steps $\Delta t_i$ that satisfy conditions of physical and numerical stability for different model modules $i = 1,2,...N$. To illustrate this point, in the CAM5 climate model convergence tests [46] for $\Delta t =1800$ s and $N = 9$, the ratios of time sizes to $\Delta t$ of radiation, deep convection, shallow convection, stratiform cloud macrophysics, stratiform cloud microphysics, vertical remapping, adiabatic fluid dynamics, resolved-scale tracer transport, and explicit numerical diffusion were 2, 1, 1, 1, 1, 1/2, 1/6, 1/6, and 1/18, respectively. 2. *Time-scale separation*. The state of a real system is often determined by more than a few processes that may operate at different time scales. The difficult task is to recognize the sizes of these time scales, at least for main processes. An additional complication is that interactions change over time—an issue rarely considered in modeling. An illustrative example is when preys change their behavior to avoid predators at a small or large time scale. One modeling strategy is to introduce a simplification by assuming that all interactions are constant (this can also lead to false estimates). The mathematical background for the time-scale separation is Tychonoff's theorem [47] stating that if fast and slow components become more separated when a small parameter approaches zero, then the arbitrary product of compact spaces is compact [48]. Fast and slow components are the categorization of processes and reactions that operate at unequal time scales. When we apply the quasi-steady-state approximation to reduce the number of variables and parameters in the model, it disables the modeling of fast elementary reactions explicitly (note that this idea was introduced by Michaelis and Menten [49]). Their influence is expressed through the parametrization by nonelementary reaction-rate functions (which is similar to the parametrization in climate models). Because of complex interactions between fast and slow components, this times-scale separation cannot be used for modeling biological systems [50]. Figure 3.2 visualizes particle diffusion paired up with the competitive birth-death interaction as an example of the time-scale separation.

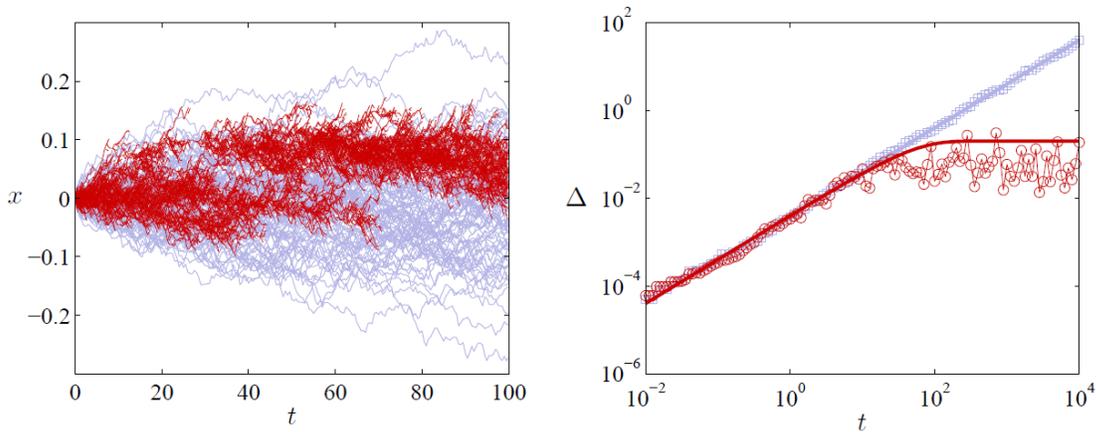

Figure 3.3. Simulation of competition-limited diffusion (*dark red*) contrasted with a collection of $N$ independent Brownian particles (*light purple*). Figure shows the particle trajectories ($x$) (*left*) and the mean square distance between pairs of particles ($\Delta$) (*right*); $t$ is the time. (Reproduced by permission from [50].)

3. *Time in functional systems* can be treated in another way. We can change our perspective of time and leave the position of external objective observers by looking at the process itself [51]–[52]. Because functional systems are regarded as languages that describe the functioning of complex systems, we call them complex systems. In such systems, the structure of time is different from o'clock time and can be considered as *functional time.* Functional time is derived from real systems where it depends ultimately on quality changes within systems and processes of their formation. The variation of the system stability influenced by these changes is considered either through its structural stability or synchronization established by measurement or through the methods of nonlinear dynamics. As an illustration, we show one simulation of the functional time formation in the process of biochemical substance exchange between cells (inter-ring arrangement). This time formation is modeled by the system of coupled difference equations [2], [52], and its model has the following features: a. Dependence of the signal strength on distances between cells is computed in the $\pi$-cell coordinate system. b. Stability of the system is synchronized globally only when the largest Lyapunov exponent of the system is negative (coded 1, 0 otherwise). c. Signals coming from the system to the location of an observer are indicators that the system does not lose its functionality; that functionality can be terminated if so-called cardinal components lose their functionality. If the system does not send any signal from non-cardinal components, it means that it sends signals but not the signal from that component or that a non-cardinal component is on a standby for synchronization with other components to send the signal. Since the coding of the state of a complex system is binary (i.e., 0 and 1 meaning either unsynchronized or synchronized), it is possible to establish functional time in the process of the exchange of biochemical substances in a multicellular system. This code is called a *functional time* *barcode* [2] and shows the "history" of the states of the system when the system evolves in time (fig. 3.3).

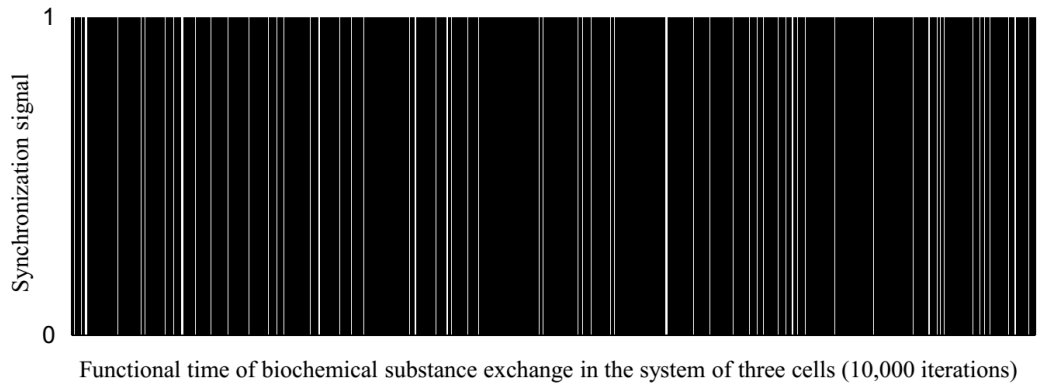

Figure 3.4. Functional time barcode of biochemical substance exchange inter-ring arrangement of cells system. (Reproduced by permission from [52].)

# Chapter 4

## Are model and theory synonymous in physics? Between epistemology and practice

### 4.1. Some background concepts and epistemology

It seems that views on the relationship between theory and model in science, art, and everyday communication can be summarized as follows: theory and model are two closely related terms, but there is a fundamental difference between them. Theory is a conceptualized framework for a phenomenon, while a model is a physical representation of a concept to make this concept more understandable. Thus, theory provides answers to various problems established in science, and models can be considered as representations created for the purpose of explaining theory. Our intention is to define the relationship between theory and model in the physics of complex systems. Perhaps the best starting point is Pierre Duhem's book *The Aim and Structure of Physical Theory* [1] written by a solid scientist and connoisseur of the philosophy of physics more than a century ago. His opinion on physical theories is condensed in the following conclusion [1]: "A physical theory is not an explanation; it is a system of mathematical propositions whose aim is to represent as simply, as completely, and as exactly as possible a whole group of experimental laws." Duhem adheres resolutely to the separation of physics from metaphysics, which shows our inability to reach the deeper levels of reality. According to him, only physical theories based on the relations of an algebraic nature (algebraic can probably be read as meaning "mathematical") between phenomena can exist. This approach is supported by many physicists (including quantum physicists) but is also considered narrow by some of them because it excludes what is necessary for the further progress of physics: the complex structure of real problems and phenomena. Duhem is strongly opposed to mechanical models: "Those who are commissioned to teach engineering are therefore eager to adopt the English methods and teach this sort of physics, which sees even in mathematical formulas nothing but models" [1]. Analyzing Duhem's understanding of theory, one gets the impression that theory and model can be used as synonyms although he does not mention the word model that is, in our opinion, incorporated in the above-mentioned statement. Duhem's understanding of *theory* and *model* allows this set of mathematical formulations (i.e., model) to be simpler than others for the description of a certain phenomenon. For example, the propagation of the temperature wave in a porous material is described by the partial differential equation $C(\partial T/\partial t) = (\partial/\partial z)(\lambda \partial T/\partial z)$, where $C$ is the volume specific weight of the porous material, and $\lambda$ is its thermal conductivity. The assumption that the thermal diffusivity $k = \lambda/C$ is constant leads to the equation for the temperature diffusion $\partial T/\partial t = k\partial^2 T/\partial z^2$ that is easier to solve in climate models. Although this assumption is physically unrealistic, it gives "reasonable" results accepted by climate modeling scientists. In other words, "Essentially all models are wrong, but some are useful" [2] as George E.P. Box, a British statistician, observes.

While theory (i.e., model) *describes* phenomena in physics, the concept of the model is based on *interpretation* in mathematics. Figure 4.1 shows that if interpretation makes something true, then we say that $I$ is a model of $S$. In mathematics, the connection between formal theories and their models is the matter of model theory that is the part of mathematical logic. In the last half of the twentieth century, model theory was linked to philosophy through its discipline the philosophy of mathematical practice.

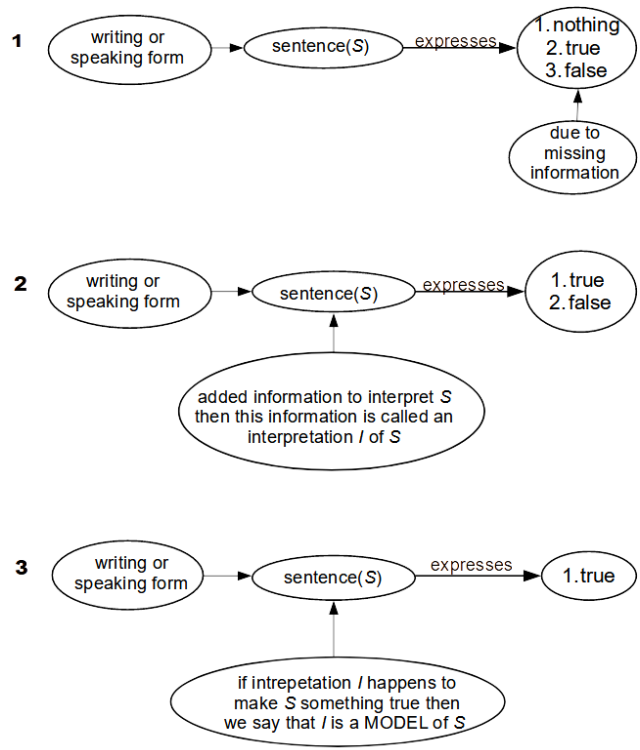

Figure 4.1. Flow diagram of the model definition in mathematics.

It is obvious that mathematics is *conditio sine qua non* for the establishment of theory that is scientifically acceptable. Physicists have recognized similarities between some mathematical relations and relations they have observed while researching physical phenomena. Mathematical physics deals with the development of mathematical methods for physical problems (this have also improved some mathematical methods). The contribution of mathematics to physical theories is impressive: (1) *Newtonian gravity* (general relativity), (2) *classical electricity and magnetism (i.e., classical field theory)* (special relativity), (3) *classical mechanics* (special relativity, quantum mechanics), (4) *quantum mechanics* (classical mechanics, quantum field theory), (5) *general relativity* (special relativity, quantum field theory, "string theory"), (6) *quantum field theory* (classical mechanics, classical field theory, quantum field theory, "string theory"), (7) *special relativity* (classical mechanics, classical electricity and magnetism, quantum field theory, general relativity), and (8) *string theory* (quantum field theory, general relativity). String theory that studies the idea of replacing all particles of matter and force with just one element—tiny vibrating strings—has been evolving towards its explosion for the last fifty years [3]. Although without impressive progress, this theory has left a profound impact on the relationship between mathematics and physics even though some physicists disagree with it. Scientists have broadly pointed out that complex systems modeling in physics needs new mathematical approaches, or "new mathematics." It seems that the part of that mathematics is also "physical mathematics" that has been changing into the separate unit slowly driven by problems in physics, quantum gravity, string theory, and supersymmetry.

A complex system implicitly contains 11 properties although a complete definition of any complex system may not be derived. Then the question that arises is how to model a complex system. One possible answer can be found on the site of Department of Computer Sciences (The University of Sheffield) [4]: "Complex systems modelling is defined by the application of diverse mathematical, statistical and computational techniques, to generate insight into how some of the most complicated physical and natural systems in the world function." In our opinion, this definition is rather literal compared with the definition that has a fundamental meaning. In fact, we do not have an exact answer to how model is defined in the physics of complex systems since this answer maybe does not exist if we consider Gödel's incompleteness theorems and Barrow's comment in subchapter 1.4. However, what is certain is that there are obstacles to the modeling of physical systems. Namely, the number of *strategies* for a physical system is undoubtedly limited by physical laws and the fact that the concept of adaptation does not exist within that system ("tyranny" of physical laws as Stephens notes [5]).

**4.2. Choice in model building**

If we consider model choice and building from Duhem's perspective, we can say that the focus is to bring choice and building as close as possible to the theory's description. Certain questions in complex systems modeling in physics cannot be answered because they are difficult to answer or *per definitionem* have no answer. We discuss some of them, including questions that are inherent to physical complex systems. Let us begin with general principles. One of the first decisions to be made is whether the model should be *macroscopic* or *microscopic*. If we analyze a phenomenon occurring at a microscopic level, the model needs to be microscopic. If the phenomenon is macroscopic, the model can be either macroscopic or microscopic depending on the phenomenon. Closely related to this division is the one into *continuous* and *discrete* models. Macroscopic models that use basic equations to capture main microstructural features are always continuous, while microscopic models are discrete. A discrete model is transformed into a continuous one after averaging (i.e., a statistical treatment) that is the basis of classical electrodynamics. Likewise, solid state physics leads to the physics of crystals. Ultimately, one of the objectives in physics is the determination of macroscopic properties from microscopic models. Next important point is that the *symmetry* must be included in the model. For example, the symmetry of a crystal produces a lot of information about physical properties, so it must be integrated from the beginning. In the study of phase transitions, an important approach is to construct the free energy of the system (a scalar) from the components of order parameter (i.e., vectors— magnetization, polarization, etc.) by using the symmetries of the material. The problem becomes more complicated when we construct models based on hypotheses about certain particles or field quanta. Usually, we do not know types of interactions between involved objects, so we base our idea on symmetries. We start with the symmetries of the free space-homogeneity and isotropy and with time—homogeneity. Time anisotropy (i.e., the existence of a privileged direction—the arrow of time) is discussed in detail in subchapter 3.1. To relate our model to experimental results, we introduce some abstract symmetry properties and require that the model satisfies those assumptions (quantum field theory is an example). It is common practice for physicists to build the model in two steps. We first consider a system with its intrinsic symmetries and then examine interactions between its constituents or with an external agent that possess its own symmetry. The interesting effects that appear in that moment are well described with the expression "C'est la dissymétrie qui cree le phénomène" (meaning "it is the asymmetry that creates a phenomenon") written by Pierre Curie in his paper [6] in 1894.

One very important information is the order of the magnitude of quantities. Suppose that we know, from either experimental data or general principles, that some quantities have larger values than others (this comparison is only correct if all quantities are expressed in dimensionless units). In that case, a small parameter takes on a value of the inverse of the largest quantity. Once we include a small parameter in the model, there are several ways to simplify calculations. The most widely used method for this purpose is the theory of perturbations in which a quantity is represented as a power series in terms of a small parameter although its convergence is often questionable. Further, involved approximations are sometimes treated as different models. Finally, we often encounter the problem of the analytical expression for the dependence between two quantities. The most suitable empirical guess for fitting data is *the power law*—that is, one quantity (variable) can be represented as a certain power of the other quantity (variable). The power is a number, positive or negative, integer, or fraction; most empirical models are based on it.

Complex systems may be defined as those that have many degrees of freedom and nonlinear interactions. Still, these conditions are satisfied by all phenomena in nature, so they are not sufficient assumptions for analyzing complex systems. For example, it is accepted that the edge of chaos is a transition space between order and disorder. It exists within various complex systems, but Stephens [5] says that the edge of chaos cannot differentiate between complex and noncomplex forms. Therefore, it is a characteristic of complex systems but not a defining one. Although experience tells us so, not everything is covered by the power law in nature and physical world. There are various examples of effective degrees of freedom with characteristic scales—objects in a macroscopic world that range from massive elementary particles to macromolecules. However, the fact that effective degrees of freedom differ significantly between characteristic scales influences the choice of the model. In addition, such effective degrees of freedom show a hierarchical connection.

Do we model complexity or complex system? This can be more understandable if we consider the illustration of a simple mathematical model [7] for a group of point particles

$$d_i = -\sum_{j \neq i} \frac{c_j(t) - c_i(t)}{|(c_j(t) - c_i(t)|} + \sum_{j=1}^{n} \frac{v_j(t)}{|v_j(t)|}, \qquad (4.2.1)$$

where $c_i$ are position vectors, $v_i$ are velocities, and $\hat{d}_i = d_i/|d_i|$ is the direction of force that occurs because of the interaction of $i$-th particle with other particles. The terms on the right-hand side are the repulsion and attraction between particles, respectively. The force that aligns the direction of a particle with $d_i$ is generated stochastically by adding a small random number to it. This simple model for point particles cannot separate complex from noncomplex systems. There is nothing to suggest, such as the hierarchy of effective degrees of freedom or a strategy, a more meaningful approach to describing an element of a complex system. This model is applied effectively to modeling the dynamics of fish shoal [7] that can be, without any doubt, attributed to complex systems. It is natural to ask ourselves how a simple equation can model a complex system. And why do we model a complex system but not its most discriminating feature—complexity [5]?

The relationship between complex systems and the host environment is crucial for the choice of the model and its building. The internal state of the system can change dynamically influenced by changes in the surrounding environment [8]. The number of different strategies determines dynamic rules for updating the status of the system. This is an important characteristic of biological complex systems (meaning that they can adapt to a new updated rule), or the strategy within a hostile environment. What is the situation in physics with respect to this strategy? There are two problems: 1. The correspondence between the surrounding environment and model. 2. Each level of describing complex systems may be autonomous without any connection with other levels. Let us consider molecules and atoms with a low ionization potential in the contact with a heated surface. Thermal conditions cause molecules to dissolve; thus, this environment is not suitable for molecules, so they cannot survive nor adapt to it. Similarly, atoms start to dissociate by losing electrons when we rise the temperature. In conclusion, it seems that physics is forever getting caught between microscopic and macroscopic states. The deeper understanding of a complex physical phenomenon can provide a new update of internal rules and, as such, be the further strategy in physics.

## 4.3. Discrete versus continuous dichotomy: time and space in model building

There is a deep interrelation between discrete and continuous behavior and the corresponding strategies for building the complex systems models in physics. Annick Lesne [10] offers comprehensive but concise comments on "the inevitable dance" between them: "A key point is that *the discrete is not an approximation of the continuum nor the converse*. Great care should be taken when passing from discrete to continuous models and conversely, since their natures are irremediably different. Paradoxes and inconsistencies between discrete and continuous viewpoints only appear when we forget that our descriptions, and even physical laws, are only *idealized abstractions, tangent to reality in an appropriate scale range*, and unavoidably bounded above and below." Both exist in any physical phenomenon and signify a fundamental dichotomy in mathematics [9]. If we take into consideration the connotation of continuous in physics and mathematics, then it is less equivocal in physics because it relates only to the time and smoothness of *continuous dynamical systems*. In physics, this relation occurs in time and space: 1. *Discrete versus continuous in time*. We first look at the relationships between discrete and continuous time and try to bridge their differences that occur in theoretical modeling, data analysis, and numerical implementation. One of the discretization procedures is the *Poincare section discretization method* [11]. If $S$ is an autonomous differentiable dynamical system generating a flow with a periodic orbit $O$ of period $T$, then the successive intersections $x_0, x_1, x_2$ ... of a continuous trajectory with the hypersurface $X$ define the return map $\varphi_X$ (*Poincare map*), where $\varphi_X(x_0) = x_1$, $\varphi_X(x_1) = x_2$, and so forth. The time when the intersection occurs is called the return time [12], and the discretization step is not chosen arbitrarily since it is already specified by dynamics. In addition to *Poincare map*, there are also: a. The *Birkhoff map* [13] in which the knowledge of the pre-collision state determines the dynamic recursion. b. *Discrete models* that can be introduced directly and are used in population dynamics. c. *Euler's discreditation scheme* that is widely popular in physics. The theory of dynamical systems in physics represents a mathematical basis for the time evolution of a phenomena. The system is given in terms of physical laws yielding an evolution equation whose parameters describe the current state of the surrounding environment that may change with time in autonomous dynamical systems. A difference equation $x_{n+1} = f(x_n), n \in \mathbb{Z}^+$, where $f: \mathbb{R}^d \to \mathbb{R}^d$ is a first-order autonomous difference equation on the state space $\mathbb{R}^d$ ($\mathbb{Z}^+$ denotes the nonnegative integers). There is no loss of generality in the restriction to the first-order equations since higher-order difference equations can be reformulated by using an appropriate higher dimensional state space. The successive iteration of an autonomous difference equation generates the forward solution mapping $\pi: \mathbb{Z}^+ \times \mathbb{R}^d \to \mathbb{R}^d$, $x_n = \pi(n, x_0) = f^n(x_0) = \underbrace{f \circ f \circ \cdots \circ f(x_0)}_{n \text{ times}}$ that satisfies the initial condition $\pi(x_0) = x_0$ and the semigroup property $x_n = \pi(n, \pi(m, x_0)) = f^n(\pi(m, x_0))) = f^n \circ f^m = f^{n+m}(x_0) = \pi(m + n, x_0)$ for all $m, n \in \mathbb{Z}^+, x_0 \in \mathbb{R}^d$. Equation $x_{n+1} = f(x_n)$ can be solved numerically by using the known initial condition to step either forward $x_{n+1} = x_n + \Delta\tau f(x_n)$ or backward $x_{n+1} = x_n + (\Delta\tau/(1 - \Delta\tau \frac{\partial f}{\partial x}(x_n)))f(x_n)$ in time, where $\Delta\tau$ is the time step, and $n$ is the time iteration. 2. *Discrete versus continuous in real space*. The three-dimensional ($d = 3$) space (or higher dimension) may be seen either as a continuum in the Euclidian space filled with geometric objects of the same dimension that determine the position or as a *lattice* represented with a series of points arranged in a different pattern, where position is characterized by $d$ integers. Lattice models are effective and reliable for the analysis of the universal properties of physical systems having the same symmetries and geometrical properties. To see an object in a macro or micro world, for one is necessary to have appropriate

"optics" through which the observed object seems to be intrinsically discrete or even isolated (individual atoms can be seen under the UK's super STEM electron microscope). Taking as an example, the atom at small scales (those of quantum mechanics) is delocalized. Because the position of a particle is determined by a probability distribution, we cannot say that discrete objects offer more objective perspective than an arbitral division of space into cells [10]. 3. *Discrete versus continuous in phase space*. The system state can be represented by a continuum (vector space) or varies within a finite or countable set of configurations. The relationship between agent-based and kinetic continuous state descriptions have taken place in complex systems modeling: population dynamics, granular media, chemical kinetics at different scales, etc. 4. *Discrete versus continuous in conjugate space*. This is considered in the context of spectral analyses since spectra offer another modality to a dichotomy. In spectral analyses, the underlaying real space is split into cells (conjugate space), so they are powerful methods for decomposing physical behavior into elementary components. This method is mainly applied when we want to have a richer picture of the system's behavior. Spectra include frequencies (or time periods), wave vectors (or wavelengths), energy levels, correlation times, and amplification rates.

We close this subchapter with some comments on digital computing. As a matter of fact, this process is bounded by an accuracy limitation. There is a contrary variation between the *method error* (i.e., the discretization error) and the *round-off error* against the step size whose chosen value regulates 1. model stability as well as its efficiency; 2. computational time; thus, a *computational uncertainty principle* is similar to the well-known Heisenberg uncertainty relation (position and momentum). The mathematical expressions of the principle are $\Delta e + \Delta \tilde{r} \geq C$ and (2) $\Delta e \cdot \Delta \tilde{r} \geq \sigma$, where $\Delta e$ is a measure of uncertainty attributed to the limitation of a numerical method itself, $\tilde{r}$ is a measure of uncertainty resulted from the limitation of computer accuracy, $C$ and $\sigma$ are positive numbers dependent on differential equations, and the machine precision is finite [14]. If we relate this principle to discretization and the round-off error, it means that if one error is smaller than the other one, then the adjoint variable is greater. The choice of the step size is of crucial importance for numerical computations, especially in their uses for solving partial differential equations. Certainly, the numerical method places upper limits on the step size, while the number of integration steps is limited by finite computer accuracy.

**4.4. Predictability of complex systems. Lyapunov and Kolmogorov times**

Predictability of a complex system usually refers to 1. the system time evolution from which we can get information; 2. the content of obtained information. When predictability is considered, our attention is mostly on a macroscopic model that predicts the state of the system for a longer period of time and spatial scale. If we evaluate how much the final state depends on the content of information in the initial state and history of the system, we speak about predictability in the context of a microscopic model.

A complex system in physics includes numerous elements that are important but not necessarily sufficient for the functioning of the system (see subchapter 4.3). The questions are what it means to model a physical complex system and whether it is possible to get information from that model. It appears that there are no simple and straightforward answers to these questions. Some features of components in complex systems cannot be measured successfully. This complicates model building, and even simple models produce different results. Physicists often use shorter computational methods to obtain results, and this practice is only possible when 1. computations are more sophisticated than those performed by physical systems; 2. a physical system behaves like a computer; this is known as *computational irreducibility* introduced by Wolfram [15]. It should be noted that the system's behavior can only be found by a direct simulation or observation; predictability based on a general procedure is not possible. For example, computationally irreducible physical processes and even computationally reducible ones at a coarse-grained level of description can be predictable [16]. Predictability of disordered spin systems that follow a deep quench is a complex issue. Long-term predictability in the mean-field models fully determines the final state. In contrast, predictability can seriously be reduced in short-range models [17].

The greatest and most visible breakthrough in physical complex systems modeling was made in the interaction between the earth and the atmosphere in meteorology. These models, at different spatial and temporal scales, have opened up new horizons not only in atmospheric physics but also in other sciences; therefore, there is a huge interest for their improvements to obtain better predictability. Edward Lorenz [18] analyzes the outcomes of the integration of a set of 12 ordinary differential equations from a simplified two-layer baroclinic model of the atmosphere. He concludes that the solutions of the model are not necessarily stationary nor periodic (this was a prevailing opinion at the time). Furthermore, he obtains the solutions that vary like weather patterns without any evidence of periodicity. Lorenz establishes the theoretical basis of weather and climate predictability (in Lorenz's sense) in the theory of error growth [19]. For a system of dimension $N$, if $x(t)$ is a small error, then the error growth from $t_0$ to $t_1$ is given by $x(t) = Z(t_0, t_1)x(t_0)$, where $Z$ is a square matrix depending on nonlinear equations in a model between $t_0$ and $t_1$. Supplementary, it was shown that the values of errors lie inside a small sphere of radius $\epsilon$ that evolve into an ellipsoid whose semiaxes are $\epsilon\delta_i$, where $\delta_1, \ldots, \delta_N$ are the singular values of $E$ or square root of the eigenvalues of $EE^T$. If the singular values are arranged in a decreasing order, the error growth occurs if $\delta_1 > 1$. Eigenvectors of $EE^T$ represent the orientation of errors.

During a certain interval of time, errors grow or decay depending on the magnitude of $\delta_i$. The overall growth of small errors can be determined by considering the interval $(t - t_0)$ to be large. The limiting values defined as $\lambda_i = (1/(t - t_0)) \lim_{t \to \infty} ln\delta_i$ are called the Lyapunov exponents. $\lambda_i$ are generally independent of the initial state for many systems and represent average exponential growth or decay. A similar definition for the eigenvectors of $EE^T$ in the limit of the large time interval yields the Lyapunov vectors that are used in the predictability analysis and as initial perturbations for ensemble forecasting in numerical weather prediction. In short, the status of predictability of current weather forecasting and climate models is: 1. Reliable forecasts cannot be made for more than 10 days. 2. Instantaneous states in climate models cannot be predicted, and predictions are only possible for some aspects of climate variability [20]. In addition, chaos in deterministic nonlinear systems can occur in vertical and horizontal fluxes exchange at environmental interfaces of the earth-atmosphere system [8].

We also point out one time scale called *prediction horizon*—the Lyapunov time $t_{lyap} = 1/\lambda_{max}$ (which is expressed in the units of the recorded series), where $\lambda_{max}$ is the largest positive Lyapunov exponent. It is a period after which a dynamical system becomes unpredictable and enters a chaotic state, so it indicates the limits of predictability. If $t_{lyap}$ increases when $\lambda \to 0$, then long-term accurate predictions are possible (fig. 4.2 *upper*). Research suggested that $t_{lyap}$ overestimates the actual value of the period. To correct this overestimation, Mihailović *et al.* [21] introduce the *Kolmogorov time* $t_{kol} = 1/K_c$, where $K_c$ is the Kolmogorov complexity. This time quantifies the size of the time window within which complexity remains unchanged. Hence, the presence of a narrow window $t_{kol}$ significantly reduces the length of the effective prediction horizon (fig. 4.2 *lower*).

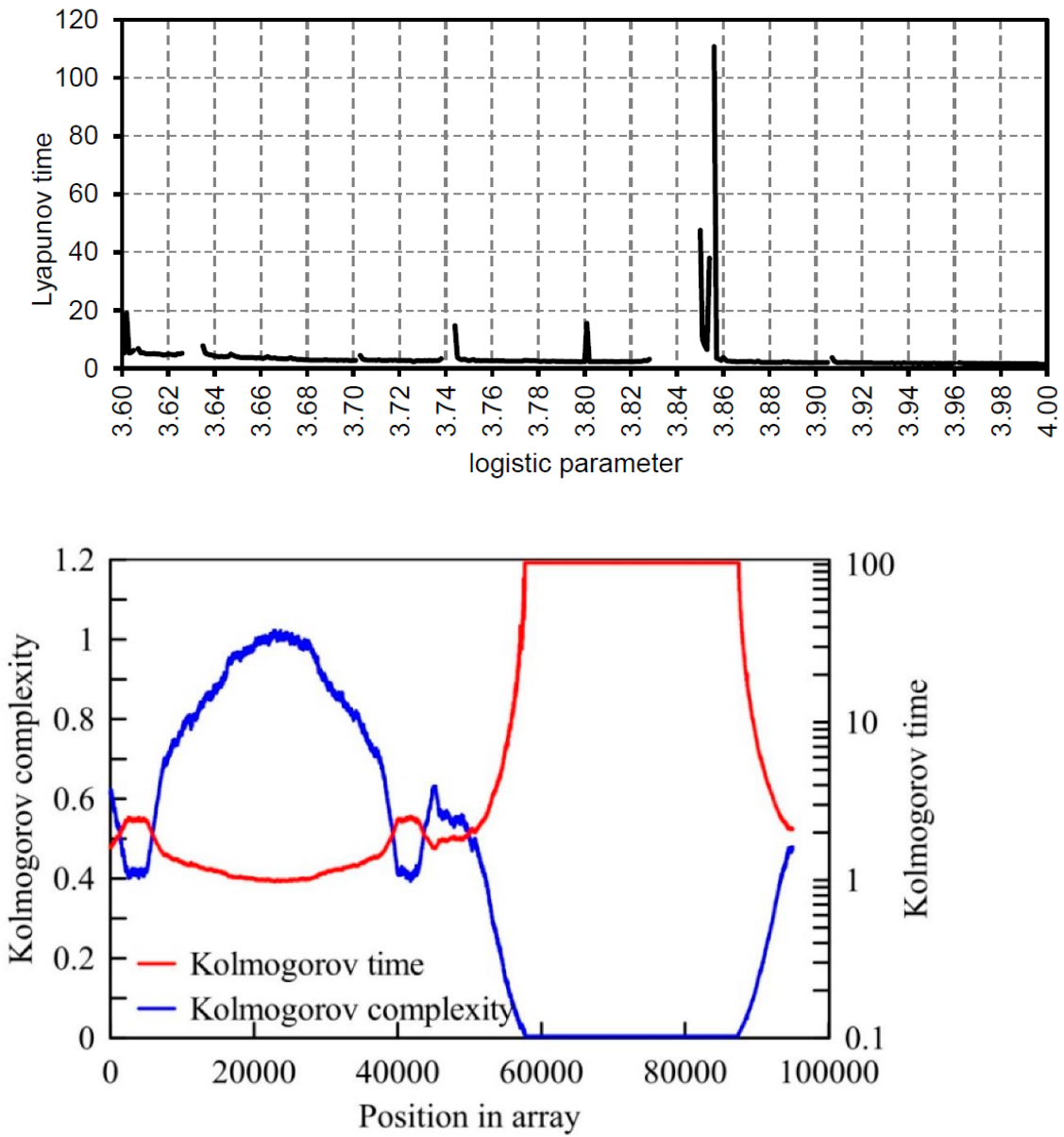

Figure 4.2. Prediction horizons for logistic equation. Lyapunov time (*upper*); empty intervals indicate $t_{lyap} < 0$. Kolmogorov time (*lower*); logistic parameter $r = 3.75 + 0.25 sin\left((2\pi)i/N\right)$, where $i$ is the position in time, $N$ is the size of time series; the running Kolmogorov complexity

$K_c$ was calculated using the window of 5000 samples long. (Reproduced by permission from Mihailović *et al.* [21].)

### 4.5. Chaos at environmental interfaces in climate models

From the very beginning of the work on climate models, there have been many debates about what models should include, how reliable they are, and issues related to chaos. Opinions on the complexity of climate models are different and narrow because extravagance and way they are presented distract the scientific community from more fundamental issues. In this subchapter, we consider climate models in the context of their modeling and as highly complex systems that consist of several major components: the atmosphere, the hydrosphere, the cryosphere, the land surface with the biosphere, and the interfacial fluxes between these components. According to observations, the climate system exhibits chaotic behavior at all time scales [22].

> Finite systems of deterministic ordinary nonlinear differential equations may be designed to represent forced dissipative hydrodynamics flow. Solutions of these equations can be identified with trajectories in phase space. For those systems with bounded solutions, *it is found that non-periodic solutions are ordinarily unstable with respect to small modifications, so that slightly differing initial can evolve into considerably different states. Systems with bounded solutions are shown to possess bounded numerical solutions.* A simple system representing cellular convection is solved numerically. *All of the solutions found to be unstable, and almost all of them are nonperiodic*. The feasibility of very long-range weather prediction is examined in the light of these results [23] (italics added).

Lorenz's paper "Deterministic Nonperiodic Flow" [23] that formed the basis for chaos theory was one of the greatest achievements in physics in the twentieth century although rare scientists noticed that at the time. Lorenz delivered the lecture "Predictability: Does the Flap of a Butterfly's Wings in Brazil Set off a Tornado in Texas?" [24] at the session of the annual meeting of the American Association for the Advancement of Science in December 1972, while James Gleick's book *Chaos: Making a New Science* [25] made the term "butterfly effect" and Lorenz's discovery popular fifteen years later.

It is accepted that weather is in the state of deterministic chaos caused by nonlinearities in the Navier-Stokes equations. The system may not be predictable for more than seven days because of the sensitivity to initial conditions. Chaos in climate models occurs for the same reasons, but there is one significant difference. Its source is the linkage of different subsystems that produces something more complex compared with deterministic chaos in weather models. Weather and climate models are basically identical. Climate models function because weather is chaotic and not random. When the climate model is integrated over a longer period of time, climate emerges from chaos. Here, we discuss chaos that occurs in the process of energy exchange at *environmental interfaces* [8] between land and the atmosphere in climate models.

In climate models, the soil surface-air layer can be treated as an environmental interface (i.e., a dynamical system) that is also sensitive to initial conditions. In such a system, chaotic fluctuations occur while we calculate the environmental interface temperatures in soil surface and deeper soil layer via two partial differential equations [26]. Depending on the climate model grid cell and atmospheric conditions at a reference level, these equations can be written in the form of the two coupled difference equations [27]–[28]

$$x_{n+1} = rx_n(1 - x_n) + \varepsilon y_n, \qquad (4.5.1)$$

$$y_{n+1} = \varepsilon(x_n + y_n), \qquad (4.5.2)$$

where $x_n$ and $y_n$ are dimensionless environmental interfaces temperatures at soil surface and deeper soil layer, respectively; $r$ is the coefficient, and $\varepsilon$ is the coupling parameter, where $r \in [0,4]$ and $\varepsilon \in [0,1]$. This system is observed when 1. the model grid cell consists of the patches similar to ones showed in fig. 4.3; 2. the time step is not chosen properly because of limitation by the computational uncertainty (see subchapter 4.3).

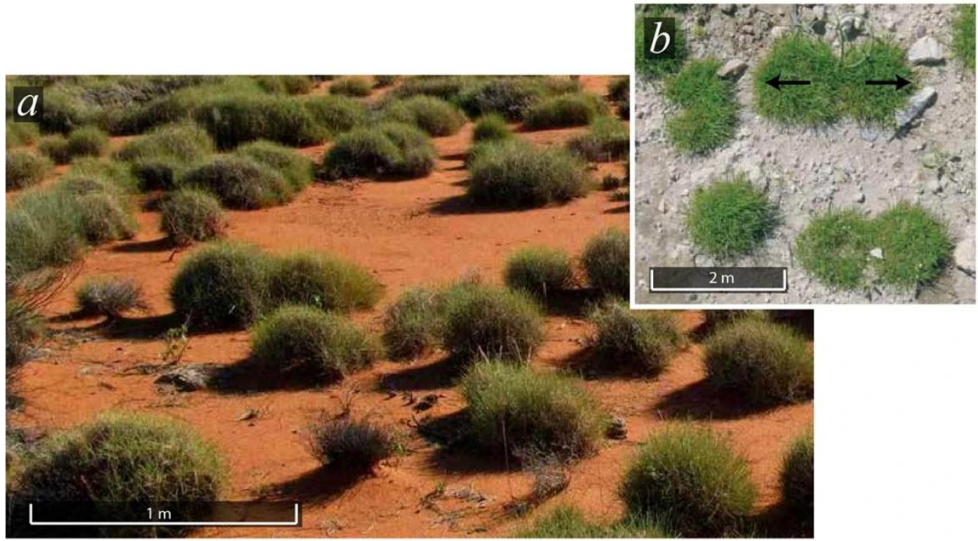

Figure 4.3. Cover type in the climate model grid cell that can determine the occurrence of chaos. (a) *Spinifex grassland*, Yakabindi station, Western Australia (courtesy of Vilis Nams, Dalhousie University, Canada). (b) Patterns of *P. bulbosa* observed in the Northern Negev. (Reproduced by permission from Meron *et al.* [29].)

The occurrence of an attractor in the system (4.5.1) is possible for the small values of the coupling parameter ε in the phase space (fig. 4.4). In this case, the system (4.5.1) is close to Hennon's map; thus, the attractor is similar to Hennon's one.

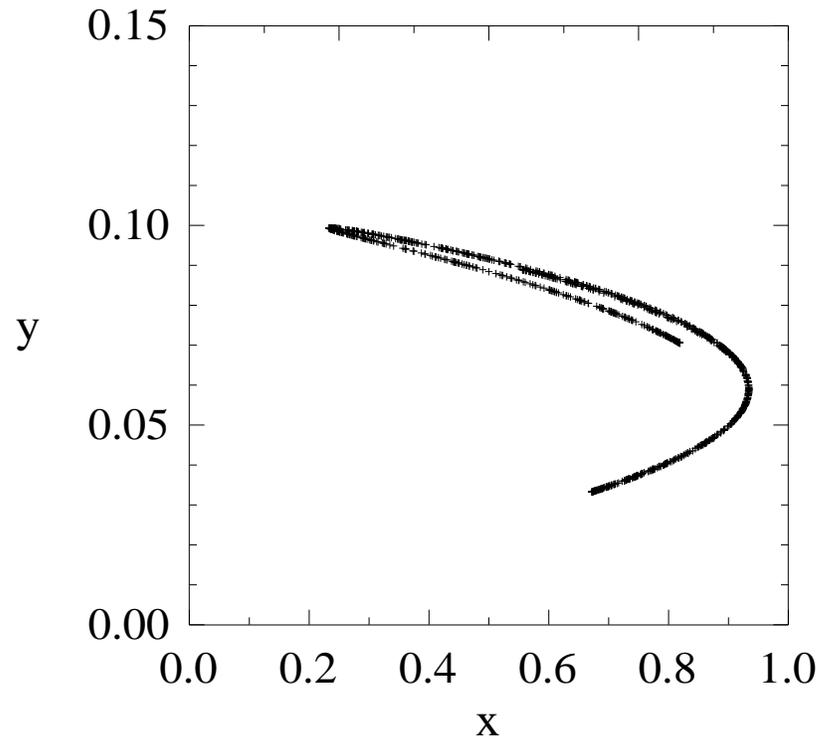

Figure 4.4. The attractor in the phase space of the coupled system (4.5.1) for $r = 0.1 = 3.7$ and $\varepsilon = 0.1$ with the initial conditions $x_0 = 0.2$ and $y_0 = 0.4$. (Reproduced by permission from Mimić [28].)

# Chapter 5

## How to assimilate hitherto inaccessible information?

**5.1. Physicality and abstractness of information. Concept of information**

Communication is the transfer of information from one point (a place, person, or group) to another. It involves at least one sender, message, and the recipient. In terms of medium, it can be divided into several distinctive categories: (1) spoken or verbal communication, (2) nonverbal communication, (3) written communication, and (4) visualization. If we look at these categories, we can ask ourselves a question: Is information physical or abstract? *Physicality* implies that information has a physical manifestation, or physical existence. Landauer [1] comes to the following conclusion: "Information is inevitably inscribed in a physical medium. It is not an abstract entity. It can be denoted by a hole in a punched card (notwithstanding this long-abandoned technology it is still noticeable as an example), by the orientation of a nuclear spin, or by the pulses transmitted by a neuron. The quaint notion that information has an existence independent of its physical manifestation is still seriously advocated. This concept, very likely, has its roots in the fact that we were aware of mental information long before we realized that it, too, utilized real physical degrees of freedom." On the contrary, abstractness assumes that information is an idea that exists in our minds. Timpson [2] argues that information is an abstract entity: "Information, what is produced by a source, or what is transmitted, is not a concrete thing or stuff. It is not so, because, as we have seen, what is produced/transmitted is a sequence type and types are abstract. They are not themselves part of the contents of the material world, nor do they have a spatiotemporal location." The important conclusion that can be drawn from this elaboration is that information can be either physical or abstract, but it certainly exists outside its physical manifestation and our ability to detect it.

We constantly respond to external stimuli by using our senses and process our inner thoughts and feelings. In fact, almost anything existing and happening in the universe can be perceived as information: "The phenomenon being observed is information, the pattern of organization of matter and energy as it exists in the universe and in living beings. The fact that we are observing, however, and claiming the objective existence of patterns of organization such as neurally stored memories, does not imply that our understanding or construction of that objective existence is true, complete, correct, or the only possible understanding. Nor does this claim imply that we deny the subjective variations and uniqueness in each individual's perception, extraction, and use of information in their minds and surroundings" [3].

Various theoretical approaches to the *concept of information* have been developed and argued in detail. They are rooted in different disciplines like physics, mathematics, logic, natural sciences, and philosophy and gather around two central properties: 1. *Information is extensive*. Extensiveness emerges from our interactions with the surrounding environment when we observe, count, and measure objects and structures. The fundamental concept is that of *additivity*: the combination of two successive and independent events. For example, let 0.8 be the probability that I will go for a walk and 0.5 the probability that I will receive a call on a cell phone (this happens independently of whether I am at home or not). Then the probability of the realization of both events is 0.40. But what about the information included in these independent events? If I am a little surprised by the event and another event happens, then my total surprise depends only on the probability of another event. In conclusion, the total information should be the sum of two independent amounts of information because the events are also independent. 2. *Information* *reduces uncertainty.* The relation between uncertainty and information was probably first characterized by English empiricists. They observed explicitly that rare events (i.e., events with low probability) were the most surprising to us and contained much information. This is formulated mathematically by Hartley [4] who defines the *amount of information $I(A)$* that we get when an element is selected from a finite set [5]. The only mathematical function that connects extensiveness and probability and defines the information in terms of the negative log of the probability $P(A)$ [5] is $I(A) := -logP(A)$ [6].

In the twentieth century, various proposals to formalize the concept of information were made through qualitative and quantitative theories of information. Some of quantitative ones include the following measures: Nyquist's function, Fisher information, the Hartley function, Shannon information, Kolmogorov complexity, entropy measures in physics (Boltzmann entropy closely related to the Hartley function), Gibbs Entropy formally equivalent to Shannon entropy, and various generalizations like Tsallis entropy and Rényi entropy. Note that these entropies are not strictly the measures of information. In physics, entropies are related to the corresponding concepts of information and quantum information, where a qubit is a generalization of the classical bit. A concise list of basic references linked to these theories is available in *The Stanford Encyclopedia of Philosophy* [5].

The connection between information in an event and probability of its occurrence entered the physical world in 1872. Ludwig Boltzmann, an Austrian physicist, invented the formula $S = klogW$ that describes the entropy $S$ of a system in terms of the logarithm of the number of possible microstates $W$ that are consistent with the observable macroscopic states of the system, where $k$ is the Boltzmann constant. From the perspective of information theory, this formula can be interpreted as a measure of the amount of entropy in the system, lack of information, or the probability of any typical microstate that is consistent with macroscopic observations.

We think that the relation $I(A)=-logP(A)$ that connects two intuitive concepts (i.e., information and probability) is fundamental like Duc de Broglie's wave-particle duality, Heisenberg uncertainty relations (position and velocity; energy and time), or the first law of thermodynamics stating that the change in internal energy of a system equals the net heat transfer into the system minus the net work done by the system.

**5.2. The metaphysics of chance (probability)**

Blaise Pascal and Pierre de Fermat are accredited as the fathers of probability since they did the fundamental work on the theory of probability by considering the gambling problem posed by French writer Antoine Gombaud, Chevalier de Méré in 1654. Probability is a concept that has contributed a lot to moving the boundaries of science (especially when it is tied with the concept of information), but it is still one of the "most puzzling" terms for which it is not possible to find the tiniest meaning. It may be understood as a more general question known as the metaphysics of chance: What makes any probability statement true or false? Various interpretations of probability have attempted to answer this question [7]. We just list some of them: classical probability, logical/evidential probability, subjective probability, frequency interpretations, propensity interpretations, and best-system interpretations.

Although often used synonymously, there is a subtle distinction between *chance* and *probability*. The metaphysical aspect of chance is inaccessible to us, that is—our experience and intuition are not sufficient for its understanding, so we want to reach that level of awareness. On the other hand, it is possible to approach that "something" but it is not completely clear how. In other words, if we cannot move towards the metaphysics of chance, we cannot move away from it, which is the exact way we understand probability.

Scientists use *probability* (which is a ratio that defines how likely an event is to happen) or *chance* (which does not have any obviousness) to grade events from rare (low probability) to frequent (high probability). In medical terminology, people increase the probability of contracting cancer by smoking. Still, there are some events that are rare, such as the landing of a randomly rolled cubical dice on one of its 12 edges. These events have very low probabilities of occurrence and are considered nearly impossible to happen. Imagine that we can throw the dice only two times, and we obtain six from the first toss and three from the second toss. We might think that this result is compatible with the dice being 1. fair—each side can occur with equal probabilities; 2. biased—some sides have a greater chance of coming up. In this example, the metaphysics of chance involves the question what makes the statement "the probability that the dice will land on its face is 1/6" true. We consider two of the above-mentioned interpretations of probability in this subchapter: the classical account of probability and the frequentist account.

According to the classical account of probability [8], probabilities or chances are abstract symmetries that could be determined *a priori* with the equal probability of occurrence. If an experiment results in $n$ equally possible outcomes, then we should assign the probability $1/n$ to each of them. So, when a coin is flipped, it can land on its head or tail. Each of these outcomes is equally possible, and the probability of both events is $1/2$. The problem with Laplace's definition can be expressed through the following question: Is it possible that the coin may land on its edge? In this case, Keynes [9] claims that the probability is not $1/2$ but $1/3$ (the head, face, and edge). He suggests that when one outcome cannot be favored over another, then they have the equal probability of occurrence. He calls that principle *the principle of indifference* and defines it as follows: "The Principle of Indifference asserts that if there is no *known* reason for predicating of our subject one rather than another of several alternatives, then relatively to such knowledge the assertions of each of these alternatives have an equal probability" [9]. However, this principle can be contradictory. Namely, we almost always obtain either the head or tail when we toss the coin, while the fact that the coin may land on its edge is completely excluded from our consciousness. Thus, the chance that this relatively simple event will happen remains unnoticed.

Frequentism, another approach, defines probability as the limit of its relative frequency in many trials so that the facts about probabilities are *a posteriori*. If we want to determine the probability that the coin will land on its head, we divide the number of times coin landed on its head by the total number of outcomes. However, there are some problems relative to frequentism. First, many events are unrepeatable, and we can obtain only one result. This eliminates other outcomes that are not the results of the experiment. In other words, instead of counting all possible outcomes, only ones that result from experiment are counted (this is known as the single case problem). Second, it is not defined why some reference classes are more suitable than others for determining the probability of the event. Here we provide an interesting probability analysis of rolling a square cuboidal dice having two squares (edge $y$ and four rectangle sides (edges $x$ and $y$) (fig. 5.1) with a focus on the probability $p$ that a square face is up after rolling the dice.

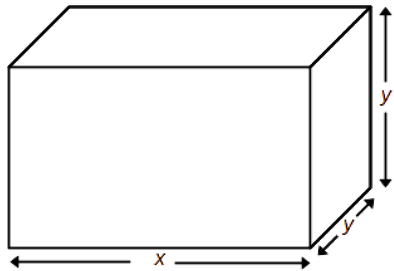

Figure 5.1. A cuboidal dice in which two opposite faces are squares of edges $y$, and the other four faces are rectangles of edges $x$ and $y$.

Mungan and Lipscombe [10] propose a model for calculating this probability by combining probabilistic arguments to produce a formula with a parameter that is fitted from results of frequency measurements of rolling the dices with different ratios $x/y$. Probabilistic arguments include the angle with respect to the ground that defines whether the dice will fall on a rectangle or square side and also the ratio of the surfaces to which the dice may fall. Both quantities depend only on the ratio $x/y$. The probability has the form

$$p = \frac{2}{2 + (x/y)^n}[1 - \frac{2}{\pi}\arctan(\frac{x}{y})^n], \qquad (5.2.1)$$

where $n$ is the parameter determined by fitting experimental data, and it takes the value between 3 and 3.5. In the particular case of $x = y$, the probability is 1/3. In the limiting case of small $x$, this can even approximate a coin.

### 5.3. Shannon information. A triangle of relationships between energy, matter, and information

In this subchapter, we describe Shannon entropy that is a foundational concept in information theory and relationship between energy, matter, and information. Claude Shannon became a scientific icon after he had established information theory in his famous paper "A Mathematical Theory of Communication" [6]. His contribution to science was described by one of the outstanding persons in information theory at the turn of twentieth century Albert Tarantola

[11]: "Shannon must rank near the top of the list of the major figures of 20th century science."

The first idea of quantifying information was proposed by Hartley [4]; however, it became clear that information could be defined and measurable after Shannon's paper had been published. Shannon puts forward the theory of converting a message generated by a source into a signal and then communicating it to a recipient over a noisy channel. Transmission is corrupted by noise, and the exact reconstruction of the message is difficult to achieve—that is, transmission cannot be done without any errors. Shannon sets the limit below which it is possible to transfer the message with zero errors. Above this limit, the communication process cannot be performed without losing information. This is known as the noisy-channel coding theorem.

Let $X$ be a discrete random variable on a finite set $\mathcal{X} = \{x_1, \ldots, x_n\}$ with a probability distribution function $p(x) = \Pr(X = x)$. *Shannon entropy* $H(X)$ of $X$ [6] is defined as

$$H(X) = -\sum_{x \in \mathcal{X}} p(x) \log_b x. \qquad (5.3.1)$$

The logarithm is usually taken to be of the base 2, in which case the entropy is measured in bits (one *bit* is the smallest unit of information having a single binary value, either 0 or 1). This entropy is the average content of information contained in the message produced by the source. While the state of a bit can only be either 0 or 1, the general state of a *qubit*, according to quantum mechanics, can be a *coherent superposition* of both [12]. A measurement of a classical bit would not disturb its state; contrary, a measurement of a qubit permanently disturbs the superposition state. Vopson [13] estimates the amount of encoded information in all the visible matter in the universe by using Shannon's information theory. He determines that each particle in the observable universe contains 1, 509 bits of information, and there are $\sim 6 \times 10^{80}$ bits of information stored in all the matter particles of the observable universe. Shannon showed that a higher entropy correlates with more information content and vice versa, or the more uncertainty is involved in transmission, the more information it holds. This is the reason why Shannon entropy is considered as a measure of uncertainty apart from being a measure of information.

Let us also mention Shannon's formula in relation to entropy in physics. Shannon called his measure "entropy" because its mathematical expression was identical to the mathematical expression of thermodynamic entropy in statistical mechanics [14]. However, those quantities are not the same since Shannon entropy can be defined for any probability distribution, while Boltzmann's entropy is related only to the specific probabilities in thermodynamics. There is an interesting story about how Shannon decided on calling his measure entropy. When he was asked what he was thinking about when his measure was finally confirmed, Shannon replied: "My greatest concern was what to call it. I thought of calling it 'information,' but the word was overly used, so I decided to call it 'uncertainty.' When I discussed it with John von Neumann, he had a better idea. Von Neumann told me, 'You should call it entropy, for two reasons. In the first place your uncertainty function has been used in statistical mechanics under that name, so it already has a name. In the second place, and more important, no one knows what entropy really is, so in a debate you will always have the advantage.' " [15].

In physics, entropy originates from the second law of thermodynamics and deals with a set of events. Shannon entropy is a measure of uncertainty in information theory; this self-entropy is associated with a single event. Therefore, information and entropy are fundamentally related since the entropy measures the average amount of self-entropies contributing to a system. It can be said that entropy is a measure of the spreading of a probability that is often referred to as disorder.

Accordingly, the second law of thermodynamics can be reinterpreted as the increasing impossibility to define precise contexts at a macroscopic level.

In 1985 Claude Shannon said: "A basic idea in information theory is that information can be treated very much like a physical quantity, such as mass or energy" [16]. What did he mean with this statement? In our opinion, he set this quotation in an ontological context, that is—he suggested that information was a real objective trait of reality like matter and energy and not merely a useful mathematical fiction that could assume different forms. In other words, the question is: Can information be reduced to matter-energy and returned to us as a single element?

Umpleby [17] comes up with a suggestion that these three concepts can physically be connected via a triangle of relationships between matter, energy, and information. Symbolically, in a MEI triangle the notations are set: (1) matter (the upper vertex M), (2) energy (the left bottom vertex E), and (3) information (the right bottom vertex I). Note that we replaced his term difference with the term information at the right bottom vertex. In Umpleby's paper, the term "difference" refers to information that is, unlike matter and energy, a function of the observer [17]. Thus, information is a function of the observer, so it is not an elementary concept. In this book, we rather use the term "information" following Szilard's consideration of a relationship between information and energy [18].

The triangle sides are represented through the following relations: 1. EM side is Einstein's relationship $E = mc^2$. 2. EI side is due to Szilard who rigorously demonstrated relations of physical entropy to information in the sense of the theory of communication. That was a breakthrough in the integration of physical and cognitive concepts. Planck's relationship $E = h\nu$ is used for this side when one photon is considered equivalent to one bit [19]. Szilard also recognized that Maxwell's famous demon required information in a sorting process on the high and low energy particles. He confirmed that the measurement of the velocity of gas molecules would produce more entropy than the sorting process would remove. Thus, this demon was successfully exorcised. 3. MI side represents the so-called *Bremermann's limit* [20], or the physical laws limit computational rate of any data processing system ("computer"). Its formula was $mc^2/v$ ~ ($m$ /gram)$10^{47}$ bits per second and later replaced with $c^5/(Gh)$ ~ $10^{43}$ bits per second, where $G$ is the gravitational constant [21]. From this triangle, we can see two outcomes: 1. "Combining" EM (the relationship between energy and matter) and EI (the relationship between energy and information), we come to the relationship between information and matter (IM), at least at the atomic level. 2. Considering all the possible carriers of information, it is seen that the relationship between matter and signal is not continuous, strongly depending on the material in which a pattern appears. Thus, up to the atomic level (where Bremermann's limit applies), we can observe a pattern or set of differences in molecules (DNA), cells (neurons), organs (the brain), groups (norms), and society (culture) [22].

Vopson [23] formulates a new principle of *mass-energy-information equivalence* (MEI triangle) showing that a bit of information has a finite and quantifiable mass while it stores information. Using the mass-energy equivalence principle and $m_{bit}c^2 = k_B T ln(2)$ (side connecting M and I triangle vertices), he finds that the mass of a bit of information is $m_{bit} = k_B T ln(2)/c^2$, where $k_B$ is Boltzmann's constant, and $T$ is temperature at which the bit of information is stored. At room temperature ($T$ = 300K), the mass of a bit of information has the value of $3.19 \times 10^{-38}$ kg. Note that the mass- energy-information equivalence principle proposed in this paper is strictly applicable only to classical digital memory states at equilibrium. Information carried by relativistic media require another treatment.

## 5.4. Rare events in complex systems—what information can be derived from them?

There is a popular quote that information is power in social life and science, which is cliché. The crucial question is how we can reach the *right* information, in particular in physics? In this attempt, we bear in mind that 1. the acceptance of information is a process including physical and cognitive concepts; 2. there are limitations because of Heisenberg's uncertainty principle in physics. Information comes from an event. In modern physics, an *event* is a physical phenomenon that occurs at a specified point in space and time (see subchapter 2.4). However, an event in particle physics refers to the outcomes 1. obtained just after interaction between subatomic particles; 2. occurring in a very short time duration; 3. that are observed in well-localized region of space. An event that occurs infrequently is called a *rare event*. Rare events have low probabilities (say, order of $10^{-3}$ or less) of the occurrence according to a probability model [24]. Usually, rare events can be of significant interest. Often, discovery of these events leads to Copernican turns in physics and, accordingly, in the physics of complex systems. How to obtain an information from such events? The computation of rare event probabilities is already challenging. Analytical solutions are not available for nontrivial problems, while standard Monte Carlo simulation is computationally ineffective. Therefore, more research efforts are focused on using set theory and developing advanced stochastic simulation methods that are more efficient. We divide the rare events into *no event registered* and *extremely rare* events.

*Extremely rare event* ("*black swan*" *event*). The term "black swan"—popularized by Nassim Nicholas Taleb—refers to unpredictable events that cause tectonic changes in nature and society at all scales, including the global scale (see subchapter 1.1). Often, these events have catastrophic consequences. In his book *The Black Swan* [25], Naib Nicolas Taleb, a promoter of this term, quotes, "First, it is an outlier, as it lies outside the realm of regular expectations, because nothing in the past can convincingly point to its possibility. Second, it carries an extreme 'impact.' Third, in spite of its outlier status, human nature makes us concoct explanations for its occurrence after the fact, making it explainable and predictable." Usually, in nature events that occur with a nonnegligible probability follow the Gaussian or normal distribution. However, "black swan" event is far away from this picture. Our intuition tells us mistakenly that "black swan" events are even rarer than they are. Hence, we neglect their occurrence by moving them towards much lower probability.

What can be "black swan" in physics? Following our comments in subchapter 1.1 regarding vertical progress, we can say that in classical physics it was the discovery of the theory of relativity and quantum physics. In mathematics, the "black swan" events like "Deus ex machina" were Gödel's incompleteness theorems and Lobachevskian geometry. There are other rare and extremely rare events occurring in many physical complex systems, mostly including geological, meteorological, and climate events: earthquakes, volcanic activity, landslides, drought, wildfires, storms, and flooding. Currently, the main task in the scientific community is to develop probabilistic and dynamical techniques to estimate the probabilities of rare events accurately having in mind that it has to be done from limited data. We list some of them: (1) the generalized extreme value method from classical extreme value theory, (2) genealogical particle analysis and the Giardina-Kurchan-Lecomte-Tailleur algorithm, and (3) brute force Monte Carlo [26].

Obviously, there was a need for a new axiomatization of probability that requires another treatment in the measurement of rare and frequent events on a deeper level. Chichilnisky [27] proposes one such axiomatization relaying on the axiom of choice that lies at the very foundation of mathematics where Gödel contributed significantly with his paper from 1940 [28]. In that paper, he proved that the axiom of choice is consistent with the axioms of von Neumann-Bernays-Gödel set theory. Chichilnisky's axiomatic approach to probability with "black swan" includes three axioms about subjective probabilities related to (1) continuity and additives, (2) unbiasing against rare events, and (3) unbiasing against frequent events. Those axioms imply that subjective probabilities are neither finitely additive nor countably additive but a combination of both. However, the finitely additive part allocates more weight to rare events than the standard distributions do.

*No-event registered*. It is possible that an event that could have happened *a priori*, did not happen *a posteriori*. In this case of the event that is not registered, parameters cannot be estimated accurately. Still, a mathematical method for obtaining the boundaries of parameters is quite possible. Zlokazov [29] offers different solutions to this problem with an emphasis on the "absence of event" model. This method was applied to the data on the synthesis of the 114th element. The element was named flerovium after the Flerov Laboratory of Nuclear Reactions of the Joint Institute for Nuclear Research (Dubna, Russian Federation) where it was discovered in 1998. Flerovium is artificial chemical element having several very short-living isotopes [30]. The signal was recorded over a time interval of 34 min; the total experiment and calibration time were about 48800 min each. All signals were recorded in the same strip detector. Calibration measurement resulted in the following frequencies: implantation of recoil nuclei = 1.3 per hour; alpha particles = 1 per hour; spontaneous imitators no division was recorded for the entire calibration time. An alternative hypothesis was that all signals were random imitators or temporary Poisson processes that are perhaps the best formalism to describe random events. They are characterized by the number of random events occurring per time period that is constant, memoryless waiting time between events (independence from prehistory), and rarity (rare or low probability events) [31]. The distribution function is

$$Q_k(t) = \frac{(lt)^k}{k!} exp(-lt), \quad t \in [0, \infty], \tag{5.4.1}$$

where $l$ is the rate parameter of the Poisson distribution; $t$ is the time. Quantity $lt$ is the mathematical expectation and, at the same time, dispersion $k(t)$ at time $t$. For signals of different types, the formula describing the probabilities of their sums taking into account their order and configuration, has the form

$$Q_{sk}(t) = p_s \prod_{i=1}^{m} Q_{ki}(t), \quad t \in [0, \infty], \tag{5.4.2}$$

where $m$ is the number of types, $p_s$ is the probability of an ordered combination of $s$-type signals and $Q_{sk}$ is the total sum of these signals. Therefore, the mathematical expectation of the number of intervals having a similar configuration of signals over time (0.48800) was zero. It could be stated that the four signals contradict the hypothesis about their background origin [31].

## 5.5. Information in complex systems

What is information in complex systems? We can define it for some systems; however, we cannot find an exact answer to this question for many of them. We can only make a list of obstacles to accessibility to information. In our opinion, there are two, maybe main, reasons for that: 1. Interactions between components of a complex system are interlocked, or one component of the system triggers or prevents the action of another one automatically. 2. Additional information that originates from the system arises, but it is not visible to the observer. This hidden information is considered as a hidden variable. If it is ignored, then an analysis of such a system is impossible. In other words, it is necessary to know how parts of a complex system interact.

What is the situation regarding information in the physics of complex systems? Seemingly, there are many systems in nature from which it is much harder to extract information than it is in physics. However, physical systems have some characteristics that are not inherent to other systems. Generally, for systems that are described algorithmically as they are in physics, we have information about 1. the evolution of the states of the components of that system; 2. the evolution of its internal states (interactions) determining the update of the states at the next time step. The list of specifics of physical complex systems, including the general characteristics of complex systems, can be summarized in the following manner: 1. Physical complex systems are often inherently nonergodic and non-Markovian. 2. Occurrence of superpositions of interactions of similar strengths. 3. Physical complex systems are often chaotic since they depend strongly on initial conditions and details of the system. 4. Physical complex system equations algorithmically describing the dynamics are often nonlinear. 5. Physical complex systems are often driven systems. 6. Physical complex systems can exhibit emergence in a simple form like the liquid phase of water, which is an emergent property of the atoms involved and their interaction [32]–[33].

In general, there are methods for getting information from complex systems that can be divided into methods for analyzing data, constructing and evaluating models, and computing complexity (time series analysis, cellular automata, agent-based models, the evaluation of complex-systems models, information theory, and ways of measuring complexity). Here, we focus only on algorithmic information theory. How can we measure the amount of information contained in an event that is given to us as an observable fact? Both Shannon information theory ("classical") and algorithmic information theory are based on the idea that this amount can be measured by the minimum number of bits needed to describe the observation. Shannon's information theory considers methods that are optimal with respect to a given probability distribution. Kolmogorov's algorithmic theory is not probabilistic in nature: the length of the *shortest* computer program that outputs a string and then terminates is defined as the amount of information in the string [34]. Therefore, complexity and information are quantified in terms of the program size. Note that Kolmogorov dealt with complexity strongly motivated by information theory and problems associated with randomness.

Let us suppose that universal Turing machine program generate output $U(p) = x$. Then Kolmogorov complexity $K_c(x)$ of a given object is defined as $K_c(x) = \min \{|p|: U(p) = x\}$ that is the size of the shortest program $p$ that generates object $x$. $K_c(x) = x + const$, while $K_c(x) \approx x$ for most objects. Since $K_c(x)$ of an arbitrary object $x$ is incomputable, it is approximated by the size of the ultimate compressed version of $x$ [35]. More precisely, a binary object $x$ is compressed, and the size of the compressed object is Kolmogorov complexity $K_c(x)$. Lempel and Ziv [35] propose an algorithm for calculating the Kolmogorov complexity of a time series (Appendix A).

Complexity has advantages over other methods that detect regular behavior: 1. Concerning statistical methods, complexity does not assume stationary probabilities. 2. Regarding nonstatistical methods extracted from the theory of nonlinear dynamical systems (Takens's theorem), complexity does not assume the existence of a low-dimensional object in phase space. Thus, complexity cannot properly be calculated; it can only be estimated. The value of $K_c(N)$ is near to zero for a periodic or regular time series and near to one for a random one, if the size of time series $N$ is large enough. For a chaotic time series, it is typically halfway between 0 and 1. For a "powerfully" random sequence of a relatively short length, $K_c(N)$ can considerably be larger than one [36]. The algorithmic complexity $K_c(x)$ is approximated by some compression

algorithms— Lempel-Ziv and its variants; the resulting number (between 0 and 1) reflects an "average" since it depends on the value that is used to binarize the series. This problem is also present in statistical randomness tests. The Kolmogorov complexity spectrum takes this problem into account [37] since it calculates the complexity taking each element of the series as a threshold. Ultimately, this relates the probability distribution with compressibility. Although it is much more time-consuming than $K_c(x)$, Kolmogorov complexity spectrum does not depend critically on the time series size—that is, it does not need millions of data to produce reliable values. Figure 3 depicts the Kolmogorov spectrum for different time series. It is very easy to identify and differentiate the curves of chaotic and quasi-periodic series from those of random series. The random variations clearly differ from each other, reflecting their probability distribution. Random series peak at the median value with respect to the probability distribution.

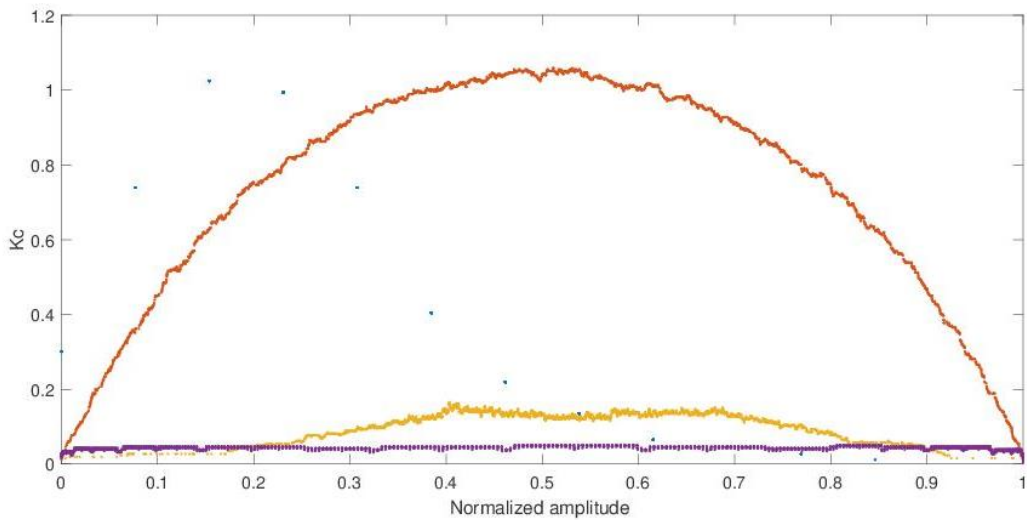

Figure 5.2. Kolmogorov spectrum (Kc) for different time series with sample sizes of 3000: (1) quasi-periodic (*violet*), (2) Lorenz attractor (*yellow*), (3) random with the constant

distribution (*brown*), and (4) random with the Poisson distribution (*blue*). (Figure courtesy of Marcelo Kovalsky.)